\documentclass{rsifpublic}

\usepackage{epstopdf}  
\usepackage{graphicx}
\usepackage{amsmath}     
\usepackage{amsfonts}    
\usepackage{amssymb}
\usepackage{color}
\usepackage{moreverb}

\begin{document}

\runningheads{R. V. Sol\'e et al.}{The ecophysics of language}

\begin{topmatter}{
\title{Diversity,  competition,  extinction:   \\  the  ecophysics  of
  language change}

\author{Ricard V.  Sol\'e$^{1,2,3}$\corrauth, Bernat Corominas-Murtra$^1$ 
\\ and Jordi Fortuny$^4$}

\address{(1) ICREA-Complex  Systems Lab, Parc de  Recerca Biom\`edica de
  Barcelona,   Universitat  Pompeu  Fabra,   Dr  Aiguader   88,  08003
  Barcelona, Spain\\ (2)  Santa Fe Institute, 1399 Hyde  Park Road, NM
  87501,  USA\\ (3)Institut de  Biologia Evolutiva.  CSIC-UPF. Passeig
  Mar\'itim de la Barceloneta, 37-49, 08003 Barcelona, Spain. \\
(4)Centre de Ling\"u\'istica Te\`orica, 
Facultat de Lletres, Edifici B, Universitat Aut\`onoma de Barcelona, 
08193 Bellaterra, Spain}

\begin{abstract}
As early indicated by Charles Darwin, languages behave and change very
much like  living species. They display  high diversity, differentiate
in space  and time, emerge  and disappear. 
A large body of literature has explored the role of information 
exchanges and communicative constraints in groups of agents 
under selective scenarios. These models have been very helpful in 
providing a rationale on how complex forms of communication emerge under 
evolutionary pressures. However, other patterns of large-scale 
organization can be described using mathematical methods ignoring 
communicative traits. These approaches consider shorter time scales 
and have been developed by exploiting both theoretical ecology and statistical
physics methods. The models are reviewed here and include extinction, 
invasion,  origination,
spatial organization,  coexistence and  diversity as key  concepts and
are very  simple in their defining  rules. Such simplicity  is used in
order to  catch the  most fundamental laws  of organization  and those
universal  ingredients   responsible  for  qualitative   traits.   The
similarities between observed and  predicted patterns indicate that an
ecological  theory   of  language   is  emerging,  supporting   (on  a
quantitative  basis) its ecological  nature, although  key differences
are  also present.   Here we  critically review  some  recent advances
lying and  outline  their  implications  and  limitations as  well  as  open
problems for future research.
\end{abstract}

\keywords{\emph{Language dynamics, extinction, diversity, competition,
    phase transitions} }}
\end{topmatter}

\corraddr{(ricard.sole@upf.edu)}

\section{Introduction}
\label{S1}

Languages  and  species  share  some remarkable  commonalities.   Such
similarities did not escape from  the attention of Charles Darwin, who
mentioned  them  a  number  of  times in  writings  and  letters  (see
Whitfield,  2008). In  {\em  The  Descent of  Man}  (Darwin, 1871)  he
explicitely says:

\begin{quote}
{\it The formation of different languages and of distinct species, and
  the proofs that both have  been developed through a gradual process,
  are curiously parallel}
\end{quote}

Languages  indeed  behave  as  some  kind  of  living  species 
(Mufwene 2001; Pagel 2009). They  exhibit a  large diversity: 
it  is estimated  that around
$6000$ different  languages exist today  in our modern  world (Krauss,
1992; Nettle and Romaine,  2000; McWorther, 2001). Languages and genes
are known to be correlated at both global (Cavalli-Sforza et al. 1988;
Cavalli-Sforza,  2000)  and  local  (see  Lansing  et  al.,  2007  and
references therein) population scales.  As it occurs with biodiversity
estimates too, the actual language diversity is unknown, and estimates
fluctuate up  to around $10000$ different  spoken languages.  Needless
to  say,  another  element  to  consider  is  the  internal  diversity
displayed  by languages themselves,  where -like  subspecies- dialects
abound.

Languages  also  display geographical  variation:  as  it occurs  with
species, they  become more  and more different  under the  presence of
physical  barriers.    They  come  to  life,  as   species  appear  by
speciation. They also get  extinct, and language extinction has become
a major problem to our cultural heritage: as it occurs with endangered
species,  many  languages  are  also  on the  verge  of  disappearance
(Crystal,  2000; Sutherland,  2003; Dalby  2003; Mufwene, 2004). 
Languages  die with
their last speaker: Crystal mentions the example of Ole Stig Andersen,
a  researcher  looking  in 1992  for  the  last  speaker of  the  West
Caucasian language Ubuh. In the words of Andersen:
\begin{quote}
{\it (The Ubuh) ... died at day break, October 8th 1992, when the last
  speaker, Tevfik Esen\c{c}, passed away.  I happened to arrive in his
  village that  very same day,  without appointment, to  interview the
  Last Speaker, only to learn that  he had died just a couple of hours
  earlier.}
\end{quote}
This  story dramatically illustrates  the last  breath of  any extinct
language. It  dies as soon  as its last  speaker dies (or  stops using
it).  It is  also interesting to observe that  the extinction risk and
its  correlation  with geographical  distribution  is  shared by  both
species and languages (Sutherland, 2003).

Language  change  involves   both  evolutionary  and  ecological  time
scales. Most theoretical studies  deal with large-scale evolution: how
languages emerge  and become shaped by natural  selection (Hawkins and
Gell-Mann 1992;  Nowak and Krakauer,  1994; Deacon 1997;  Parisi 1997;
Cangelosi  and Parisi  1998; Pinker  2000; Cangelosi  2001;  Kirby 2002; Hauser et
al.  2002; Wray 2002;  Brighton et  al. 2005;  Kosmidis et  al., 2005,
2006; Baxter  et al.,  2006; Szamado and  Szathmary 2006;  Oudeyer and
Kaplan  2007; Floreano  et al.,  2007; Lipson  2007;  Christiansen and
Chater  2008; Chater  et  al.,  2009; Nolfi  and  Mirolli 2010).   But
languages also display changes within the short time scale of one or a
few human generations.  Actually, a great deal of  what will happen to
languages in the future is  deeply related to their ecological nature.
Demographic growth, the dominant role of cities in social and economic
organization  and globalization  dynamics will  largely  shape world's
languages (Graddol, 2004).

\begin{figure*}
\centering \scalebox{0.65} {\includegraphics{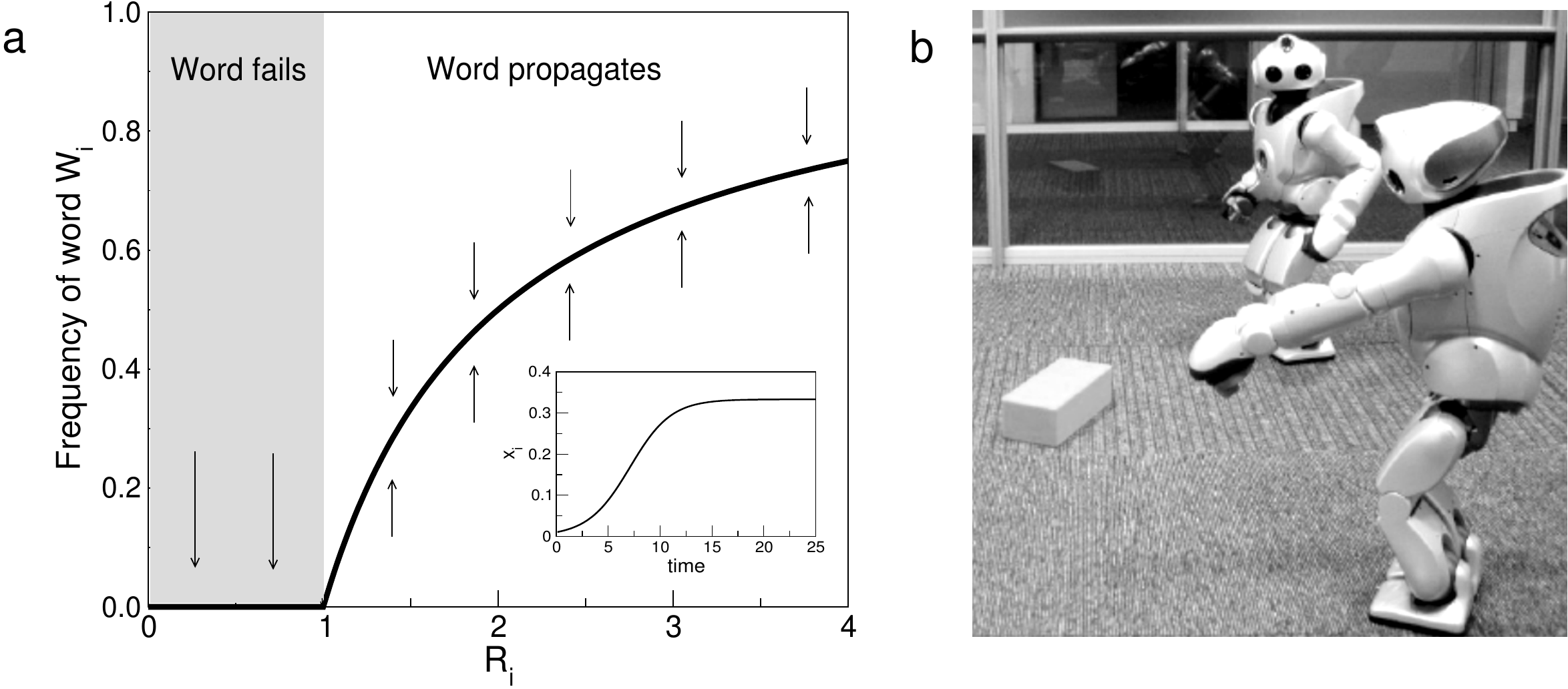}}
\caption{(a) Bifurcations  in word  learning dynamics: using  a simple
  model  of epidemic  spreading of  words, two  different  regimes are
  present. If the rate of  word learning exceeds one (i. e.  $R_i>1$),
  a stable  fraction of  the population  will use it.  If not,  then a
  well-defined threshold is found (a phase transition) leading to word
  extinction. The  inset shows an  example of the  logistic (S-shaped)
  growth  curve for  $R_i=1.5$ and  $x_i(0)=0.01$.   Lexical diffusion
  also occurs  in so called  naming games among artificial  agents (b)
  where  words are  generated, communicated  and eventually  shared by
  artificial, embodied agents such  as robots (picture courtesy of Luc
  Steels, SONY Labs).   As  common  words  get  shared, a  common  vocabulary  is
  generated and eventually stabilized. The dynamics of these exchanges
  also follows an S-shaped pattern.}
\label{epidem}
\end{figure*}

Languages evolve under centuries of accumulated modifications (this is
well illustrated by  written texts, see Howe et  al., 2001, Bennett et
al, 2003) and undergo evolutionary bursts (Atkinson et al., 2008).  On
short  time  scales they  can  be  described  in terms  of  ecological
systems.  These  rapid modifications affect  language diversity, their
internal differentiation  and even their  survival.  Different studies
using  the   perspective  of  statistical   physics  (Nettle  1999a-c;
Benedetto et al.,  2002; Stauffer and Schulze, 2005;  Wang and Minett,
2005; Ke et  al., 2002, 2008; Loreto and  Steels, 2007; Zanette, 2008;
de Oliveira et al., 2008) have been able to cope with these phenomena,
showing that  the basic trends  of language dynamics  share remarkable
similarities with the spatiotemporal behavior of complex ecosystems.

We will consider different levels of language organization, from words
to languages  as abstract entities.  The models  reviewed here explore
the  conditions  under  which   words  or  languages  can  survive  or
disappear.  The time scale is  ecological; therefore we assume that in
short time scales the dynamics of change does not affect the structure
of language  itself and thus  evolutionary models are  not considered.
Moreover,  we  do  not  intend to  quantitatively  reproduce  observed
patterns, although the predictions of the models can be tested in many
cases from  real data.  Instead, the  models we revise  try to capture
the {\em logic} of the  underlying processes in a qualitative fashion.
These models  follow the  spirit of statistical  physics in  trying to
reduce system's complexity to its bare bones.  They provide a powerful
approximation  that allow  us to  see global  patterns that  might not
depend on the intrinsic nature  of the components involved.  They also
help  highlighting  the  differences.   As will  be  discussed  below,
languages also exhibit marked departures from ecological traits.

This review critically  examines   a  set  of  models  of  increasing
complexity.  Specifically, we review recent advances within the fields
of statistical  physics and theoretical  ecology relative to  a better
understanding of language dynamics.  We begin with a very simple model
describing word propagation within a population. Next, the effects and
consequences  of competition among  linguistic variants,  with special
attention to those scenarios  leading to language extinction.  This is
expanded  by  considering   alternative  scenarios  allowing  language
coexistence  to  occur, either  through  bilingualism  or spatial  and
social  seggregation.    Although  spatial  coexistence   under  local
competition  is  shared with  ecosystems,  bilingualism  belongs to  a
different  class of  phenomenon.   All these  models  involve a  small
number of  interacting languages. The  final part of the  review deals
with  language diversity in  space and  time. Both  a simple  model of
multilingual  communities  and  available  data  on  scaling  laws  in
language diversity  are presented.  Once  again, striking similarities
and strong differences are found.  A synthesis of these ideas and open
problems  is presented  at the  end, together  with a  table comparing
language and ecosystem's properties.

\section{Lexical diffusion}

The potential set  of words used by a speakers  community is listed in
dictionaries (Miller,  1991).  They capture  a given time  snapshot of
the available vocabulary, but in reality speakers only use part of the
possible words: many are technical and thus only used by a given group
and many  are seldom used. Many  words are actually  extinct, since no
one  is  using  them.   On  the  other hand,  it  is  also  true  that
dictionaries do not  include all words used by  the community and also
that new words are likely  to be created constantly within populations
and  their origins  have  been sometimes  recorded (Chantrell,  2002).
Many  of them are  new uses  of previous  words or  recombinations and
sometimes they come from technology.  One of the challenges of current
theories of  language dynamics  is understanding how  words originate,
change  and spread  within and  between populations,  eventually being
fixed or extinct.   In this context, the appearance of  a new word has
been compared to a mutation (Cavalli-Sforza and Feldman, 1981).

As it  occurs with mutational events in  standard population genetics,
new words or sounds can disappear, randomly fluctuate or get fixed. In
this context, the idea that words, grammatical constructions or sounds
can  spread through a  given population  was originally  formulated by
William Wang.  It was  proposed in order  to explain how  {\em lexical
  diffusion}  (i.  e. the  spread  across  the  lexicon) occurs  (Wang
1969).  Such process  requires the  diffusion of  the  innovation from
speaker to speaker (Wang and Minett 2005).

\subsection{Logistic spreading}

A  very   first  modeling   approximation  to  lexical   diffusion  in
populations should account for the spread of words as a consequence of
learning  processes (Shen  1997; Wang  et al.,  2004; Wang  and Minett
2005).   Such  model  should  be  able  to  establish  the  conditions
favouring word fixation. As a  first approximation, let us assume that
each  item is incorporated  independently (Shen,  1997; Nowak  et al.,
1999).  If $x_i$ indicates the  fraction of the population knowing the
word $W_i$, the population dynamics of such word reads:
\begin{equation}
{dx_i \over dt} = R_i x_i (1-x_i) -x_i,
\end{equation}
with  $i=1, ...,  n$. The  first term  in the  right-hand side  of the
previous equation  introduces the way  words are learned.   The second
deals with deaths  of individuals at a fixed  rate (here normalized to
one). The  way words  are learned involve  a nonlinear term  where the
interactions  between  those  individuals  knowing $W_i$  (a  fraction
$x_i$) and  those ignoring  it (a fraction  $1-x_i$) are  present. The
parameter $R_i$ introduces the rate at which learning takes place.

Two   possible   equilibrium  points   are   allowed,  obtained   from
$dx_i/dt=0$. The first is $x_i^*=0$ and the second:
\begin{equation}
x_i^* = 1 - {1 \over R_i}.
\end{equation}
The first corresponds to the  extinction of $W_i$ (or its inability to
propagate)  whereas the  second involves  a stable  population knowing
$W_i$. The stability  of these fixed points is  determined by the sign
of
\begin{equation}
\lambda(x_i^*)  =  \left (  {\partial  \dot{x_i}  \over \partial  x_i}
\right )_{x_i^*}.
\end{equation}
If $\lambda(x^*)<0$ the  point is stable and will be unstable otherwise
(Kaplan and Glass, 1995; Strogatz 2001).

The larger  the value of $R_i$,  the higher the  number of individuals
using the  word. We can see  that for a  word to be maintained  in the
population  lexicon,  we  require   the  following  inequality  to  be
fulfilled:
\begin{equation}
R_i > 1.
\end{equation}
This means that  there is a threshold in the  rate of word propagation
to sustain  a stable population.  By displaying  the stable population
$x^*$  against  $R_i$ (figure  1a)  we  observe  a well-defined  phase
transition  phenomenon:  a  sharp  change  occurs  at  $R_i^c=1$,  the
critical point  separating the  two possible phases.   The subcritical
phase $R_i<1$ will inevitably lead to the loss of the word.

The  dynamical  pattern  displayed  by a  succesful  propagating  word
follows  a  so  called   $S-$shaped  curve  (see  (Niyogi,  2006)  and
references   therein concerning the
gradualness and abruptness of  linguistic change).  This can be easily
seen by  integrating the  previous model. Let  us first note  that the
original equation (1) can be re-writen as a logistic one, namely:
\begin{equation}
{dx_i \over dt} = (R_i-1) x_i \left ( 1-{x_i \over x_i^*} \right),
\end{equation}
which, for an initial condition $x_i(0)$ at $t=0$, gives a solution
\begin{equation}
x_i(t) = {x_i(0) e^{(R_i-1)t} \over 
1 + x_i(0) ( e^{(R_i-1)t} - 1) }.
\end{equation}
This  curve  is known  to  increase  exponentially  at low  population
values, describing a scenario  where words rapidly propagate, followed
by  a slow  down  as the  number  of potential  learners decays.   The
accelerated,  exponential growth  has  been dubbed  the {\em  snowball
  effect} (Wang and Minett, 2005)  and such curves have been fitted to
available  data  (Wang  1969).    Therefore,  a  central  property  of
linguistic  change,  namely its  gradualness,  can  be  derived as  an
epiphenomenon from the dynamical patterns of successful propagation in
the  case of  lexical diffusion.   A  further issue  would to  explore
whether  the gradualness  of grammatical  (phonological, morphological
and syntactical) change can be derived from equations similar to those
that model the diffusion of words.  It must be noted, from a different
perspective, that the logistic  trajectory of linguistic change may be
favored  by ``the  underlying  dynamics of  individual learners'',  as
argued by Niyogi (Niyogi 2006, p. 167).

The  previous toy  model of  word  dynamics within  populations is  an
oversimplification,  but it illustrates  fairly well  a key  aspect of
language  dynamics, which  is  also observed  in  ecology (Sol\'e  and
Bascompte,  2006):  thresholds  exist  and  play  a  role  (Nowak  and
Krakauer, 1999).   They remind us  that, beyond the gradual  nature of
change  that  we perceive  through  our  lives  (mainly affecting  the
lexicon) sudden changes are also  likely to occur. An important aspect
not  taken explicitely  into  account  by the  previous  model is  the
process  of word  generation and  modification.  Words  are originated
within populations through different  types of processes.  They become
also incorporated by invasion from foreign languages.  Once again, the
processes of  word invasion  and origination recapitulate  somehow the
mechanisms of change in biological populations.

\subsection{Multidimensional diffusion}

Several modifications  and extensions of the previous  model have been
suggested (Wang et al., 2004). They include considering multiple words
involved  in the  diffusion  process. This  scenario  would take  into
account the idea that words  interact among them in multiple ways, and
their diffusion  can be constrained or enhanced  by these interactions
(Wang and Minett 2005).  The resulting model describes the dynamics of
a  given  novelty  $x_i$  and  its  previous  form  $y_i$  (these  can
correspond  to two  word or  sounds). Assuming  conservation  of their
relative abundances, i.  e. $x_i+y_i=1$,  it is posible to show that a
set of equations
\begin{equation}
{dx_i \over dt} = (1-x_i) \left [ \alpha_{ii} x_i + 
\sum_{j \ne i}^N \alpha_{ij} x_j \right ]
\end{equation}
with  $i,j=1, ..., N$,  describes the  lexical diffusion  process. The
matrix elements  $\alpha_{ij}$ introduce  the coupling rate  between a
pair $(i,j)$ of words. It is interpreted as the rate at which adoption
of the new  word $i$ is induced by the frequency  of other novel forms
of word $j$.  As it is  formulated, the stable states are all given by
$x_i^*=1$  and   thus  (not  surprisingly)  there  is   no  place  for
extinction,  although there  exists some  evidence for  such scenario,
where new  items spread initially but eventually  decay (Ogura, 1993).
An interesting extension of this  problem could take into account both
positive and negative interactions. In this way, not only facilitation
(as given by the positive  interactions) but also competition would be
considered. In  other words,  it seems reasonable  to think  that some
words should  be incompatible with  others. This actually  matches the
problem of  species invasion and assembly  in multispecies communities
(Levins 1968;  Case 1990,  1991; Sol\'e et  al., 2002). For  an exotic
species invading  a given  community to succeed,  some community-level
constrains need  to be satisfied.  It  would be interesting  to see if
similar rules apply to the ups and downs of word spreading.

As  in the  previous  subsection, it  seems  fair to  us  to pose  the
question of  whether or not  grammatical change can be  modelled using
equations similar to those explored in the study of lexical diffusion.
As  to multidimensional  diffusion,  it may  be  worth considering  in
future research whether the diffusion  of a grammatical object such as
a  morphological paradigm or  a syntactic  structure can  be described
with an  equation analogue to  eq.  (7). It  is also worth  noting the
existence  of implicational universals  (Greenberg, 1963),  which have
the shape  {\em given  a grammatical  property x in  a language  L, we
  always  find a  property y  in L},  as well  as  the crosslinguistic
observation that  certain properties  tend to entail  other properties
{\em with overwhelmingly greater than  chance frequency}, to put it in
Greenberg's famous words.  That is, crosslinguistic grammatical change
cannot  be perfectly  mapped into  a pure  diffusion  process: certain
properties entail or tend to entail the presence or absence of certain
properties, as different words may positively or negatively interact.

\subsection{Naming games}

A related  problem which  also involves the  generation and  spread of
words  is the  so called  naming  game. The  original formulation  and
implementation of this  problem was proposed by Luc  Steels as a model
for the emergence of a shared vocabulary within a population of agents
(Steels,  2001,  2003,  2005;  see  also  Nolfi  and  Mirolli,  2010).
Originally, this approach  involved communication between two embodied
communicating agents.   These agents (figure 1b) are  able to visually
identify  objects  from their  environment,  assign  them to  randomly
generated  names  which  are  then  sent  to  the  other  agent  in  a
speaker-hearer  kind  of  interaction.   Exchanges  receive  a  payoff
everytime the same word is used by both agents to name a given object.
This is done by means of  a trial and error process where failures are
common at the beginning, as  a common emergent lexicon slowly emerges.
Specifically, the set of rules are:
\begin{enumerate}
\item
The speaker selects an object.

\item
The speaker chooses a word describing the object from its inventory of
word-object pairs.  If  it doesn't have a word then  it invents one for
the  object.   The  speaker  transmits  the word-object  pair  to  the
listener.

\item
If the  listener has the word-object  pair then the  transmission is a
success.   Both agents remove  all other  words describing  the object
from their inventory and keep only the single common word.

\item
If the listener does not  have the word-object pair, then the listener
will add  this new word  to its inventory.  And this is recorded  as a
failure.

\end{enumerate}
Eventually, a  shared, stable repertoire gets fixed.   The basic rules
can  be  easily  mapped into  a  toy  model  (the naming  game  model)
involving  many  agents,  by  using  a  statistical  physics  approach
(Baronchelli  et  al.,  2006,  2008).   Both  hardware  and  simulated
implementations display an S-shaped growth of the vocabulary, although
interesting  differences  arise  when  we take  into  account  spatial
effects and the pattern of  relations between agents, describable as a
complex network (Steels and McIntyre  2003; Dall'Asta et al., 2006; Lu
et al. 2008; Liu et al., 2009).

\section{Competition and extinction}

Languages  are  spoken by  individuals,  and  the  number of  speakers
provides a measure  of language breadth. Because of  both economic and
social factors, a given language can become more efficient than others
in recruiting  new users and  as a consequence  it can reach  a larger
fraction   or   even  exclude   the   second   language,  which   gets
extinct\footnote{Species and languages also get extinct under external
  events (such as asteroid impacts or climate change). Sudden death of
  a language  can occur due to  a volcanic eruption  killing the small
  population of speakers or (more  often) as a consequence of genocide
  (Nettle and Romaine 2002)}.  This replacement would be a consequence
of  competition, one of  the most  essential components  of ecological
dynamics, which can be applied to language dynamics too.  Early models
of  two-species competition  define  the basic  formal scenario  where
species interactions  under limited resources occur  (Case, 2000). The
standard model is provided  by the classical Lotka-Volterra equations,
namely:

\begin{figure}
\centering \scalebox{0.45} {\includegraphics{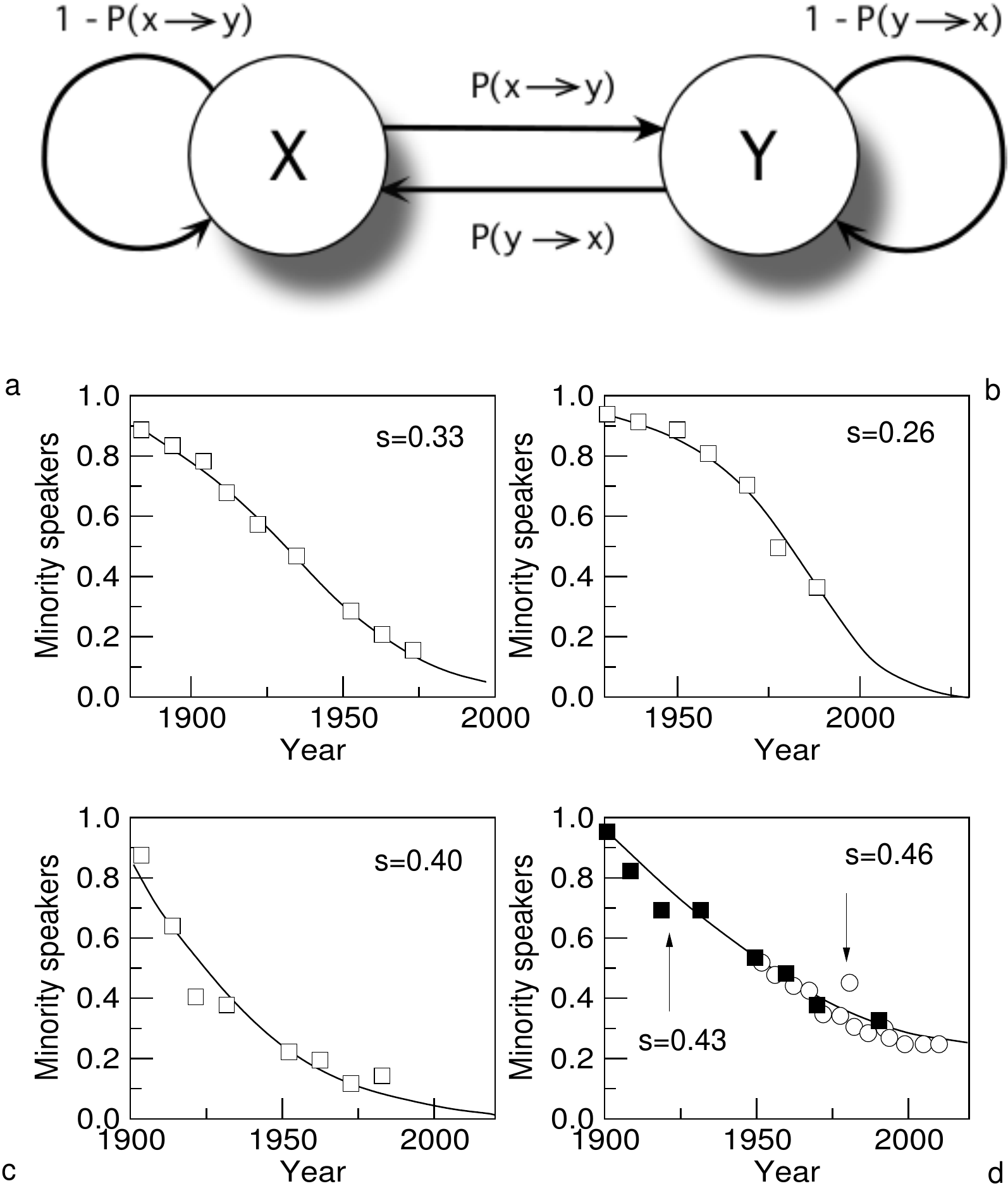}}
\caption{The dynamics of language death. Here four different cases are
  represented: (a) Scottish Gaelic,  (b) Quechua in Huanuco, Peru, (c)
  Welsh in  Monmouthshire, Wales and (d)  Welsh in all  of Wales, from
  historical data  (filled squares) and  a single modern  census (open
  circles).  Fitted curves show solutions of the Abrams-Strogatz model
  (schematically indicated in the upper plot). Redrawn from Abrams and
  Strogatz, 2003.}
\label{Strogatz}
\end{figure}

\begin{equation}
{dx \over dt} = \mu_1 x (1 - x - \beta_{12} y),
\end{equation}
\begin{equation}
{dy \over dt} = \mu_2 y (1 - y - \beta_{21} x),
\end{equation}
where $x$  and $y$ indicate the (normalized)  populations of competing
species,  $\mu_i$ indicate  their (per  capita) growth  rates  and the
coefficients $\beta_{ij}$ are the rates of interspecific competion. We
can  see that  for $\beta_{ij}=0$  two independent  logistic equations
would  be  obtained, whereas  for  non-zero  competition two  possible
scenarios are at work.

Understanding language  competition dynamics is  clearly important: if
the exclusion  scenario is  also at work,  then competition  can imply
extinction. Moreover, theoretical models can help in defining 
useful strategies for language preservation 
and revitalization (Fishman, 1991; 2001). 
Steady  language decline has been observed  in some cases,
when population records of speakers are available. This is illustrated
in figure 2, where the decay  over time of four different languages is
depicted. All  these languages  were used by  a minority  of speakers,
competing  with  a  dominant  tongue  that was  gradually  adopted  by
speakers  as  the  less  used  ones  were  abandoned.   This  type  of
increasing return is common in economics, where positive feedbacks and
amplification phenomena are common (Arthur, 1994).

A simple  model was  proposed by Abrams  and Strogatz, which  has been
shown to provide  a rationale for the shape  of language decay (Abrams
and Strogatz, 2003; Stauffer et al., 2007).  The model is based on the
assumption that two languages are  competing for a given population of
potential speakers  (the limiting resource) where we  will indicate as
$x$ and $y$  the relative frequency of each  population (assuming that
individuals are monolinguals, see  below). The dynamics is governed by
the following differential equation:
\begin{equation}
{dx \over  dt} =  y P_{\alpha,s}[y \rightarrow  x] -  x P_{\alpha,s}[x
  \rightarrow y],
\end{equation}
where it  is assumed that  $P_{\alpha,s}[x \rightarrow y]=0$  if $x=0$
and also constant  population ($x+y=1$).  The transition probabilities
depend on two parameters. The specific model reads:
\begin{equation}
{dx \over dt} = s (1-x) x^{\alpha} - (1-s)x(1-x)^{\alpha},
\end{equation}
where the $s$  parameter indicates the so called  social status of the
language.   Two  extreme equilibrium  states  are  easily found  after
imposing $dx/dt=0$.   These are $x^*=0$ (zero  population) and $x^*=1$
(all speakers use the language).  In our case, the stability criterion
gives  $\lambda(0)=s-1<0$  and  $\lambda(1)=-s<0$  and thus  both  are
stable attractors.

Together with the  exclusion points $x=0$ and $x=1$,  there is a third
equilibrium point, which can be obtained from:
\begin{equation}
s {x^*}^{\alpha -1}=(1-x^*)^{\alpha -1}(1-s),
\end{equation}
and, after some algebra one finds that:
\begin{equation}
x^*  = \left  [ 1  + \left ({s  \over 1-s}  \right )^{1 \over \alpha -1} 
  \right ]^{-1}.
\end{equation}
Given the  stable character of the  other two fixed  points, $x^*$ can
only be unstable and thus no coexistence is allowed.

The  model has  been  used to  fit  available data  on language  decay
(figure  2)  and assumes a scenario  of  minority
languages  competing with  widely  used, majority  tongues. One  clear
implication of the stability analysis is that the extinction of one of
the competing solutions is inevitable. The social parameter will influence 
which language will  get extinct.    Nonetheless,   linguistic
diversification seems  unavoidable: the language that  succeeds in the
competition situation will become more  and more diverse as it extends
through  time  and  space,  and   it  may  end  up  yielding  mutually
unintelligible linguistic variants.
\begin{figure}
\centering \scalebox{0.45} {\includegraphics{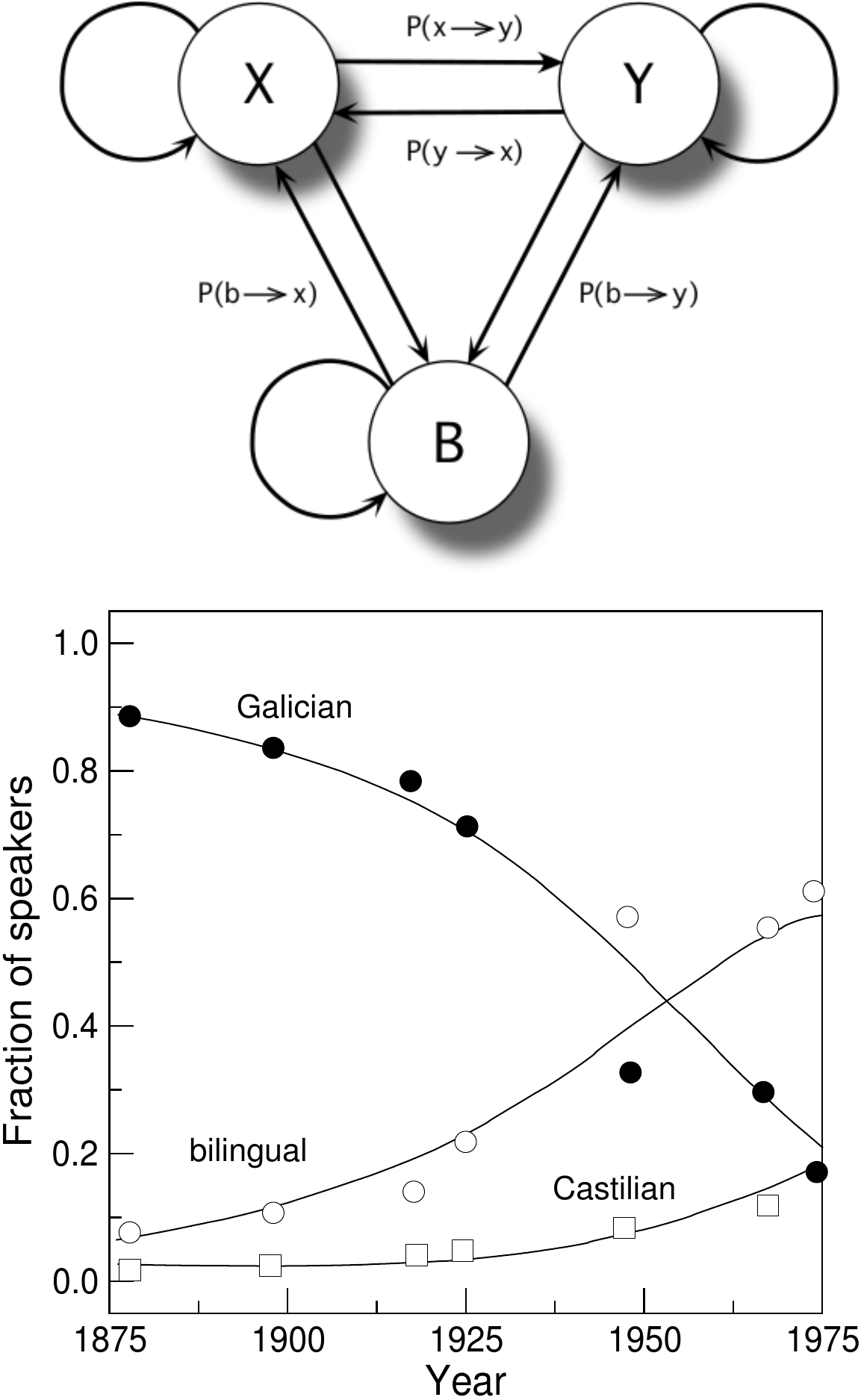}}
\caption{Dynamics  of language  use  under the  presence of  bilingual
  speakers.  Here three types of  speakers are considered (a).  In (b)
  we show the fraction of  speakers vs. time in Galicia (North Western
  Spain).  The  smooth curves (modified after Mira  and Paredes, 2005)
  are  the result  of fitting  a modified  Abrams-Strogatz  model (see
  text).}
\label{Biling}
\end{figure}

The AS model does not take into account that a fraction of individuals
is  likely (under some  circumstances) to  become bilingual.  It might
seem a  not so relevant  item, but bilingualism actually  introduces a
very  interesting ingredient  to our  view of  language change,  to be
outlined in the next section.

\section{Coexistence and bilingualism}

The  previous model is  simplified in  many respects.   By considering
human  populations as  homogeneous systems,  geographical  effects and
some idiosyncracies of human language (not shared with ecosystems) are
ignored. Spatial effects will be explored in the next section. Here we
concentrate  on a special  property of  human communities,  namely the
presence  of  individuals who  are  grammatically and  communicatively
competent on  more than one  language.  Actually, a large  fraction of
humankind uses more than one tongue for communication.  Historical
reasons  and  the influence  of  modern  invasions  by languages  like
English  makes multilingualism  an important  ingredient to  take into
account.

The  Abrams-Strogatz  model can  be  easily  expanded  (figure 3a)  by
assuming  that two languages  are present  but bilingual  speakers are
also allowed  (Mira and  Paredes, 2005; Castell\'o  et al.,  2006; see
also Minett  and Wang 2008).  The  basic idea behind  this approach is
that  the presence  of bilingual  speakers makes  language coexistence
likely to occur,  provided that the two languages  are close enough to
each other.  In this picture, three  variables are used: as  in the AS
model, $x$  and $y$ will be  the fraction of  speakers using languages
$X$ and $Y$.   Moreover, a third group $B$ using  both languages has a
size  $b$ in such  a way  that $x+y+b=1$.  Transitions are  defined in
similar ways  (figure 3a).  For  example, changes in $x$  would result
from a kinetic equation:
\begin{equation}
{dx \over dt} = y P[y \rightarrow  x]+b P[b \rightarrow x] - x \left (
P[x \rightarrow y] + P[x \rightarrow b] \right ),
\end{equation}
and the  constant population constraint  allows defining the  model in
terms of just two coupled equations, namely:
\begin{equation}
{dx  \over  dt}  =  c  \left  (  (1-x)(1-\kappa)s_x  (1-y)^{\alpha}  -
x(1-s_x)(1-x)^{\alpha} \right );
\end{equation}
\begin{equation}
{dy \over dt} = c \left ( (1-y)(1-\kappa)(1-s_x)(1-x)^{\alpha} - y s_x
(1-y)^{\alpha} \right ),
\end{equation}
where $\kappa  \in [0,1]$ is a  new parameter measuring  the degree of
similarity among  languages and the language status  are now indicated
as  $s_x$  and  $s_y=1-s_x$,  respectively.   The  $\kappa$  parameter
provides a measure of the likelihood that two single-language speakers
can communicate with each other.  It also affects the probability that
a monolingual speakers becomes bilingual. We can easily check that the
model reduces to the AS scenario for $\kappa=b=0$.

Available  data  from language  change  in  Northern  Spain (Mira  and
Paredes, 2005)  provide a test of  this model. Here  the two languages
are Castilian and Galician,  both derived from Latin.  These languages
allow a relatively good mutual understanding and parameters are easily
estimated.  For this  data set, a best fit  was obtained using $a=1.5,
s(Galician)=0.26,  c=0.1$  and  $\kappa=0.8$.   As  we  can  see,  the
apparent  decline   of  Galician  is  actually  a   consequence  of  a
simultaneous increase of Castilian monolinguals and bilinguals. 

We should be  aware of the overestimation of the  role of the $\kappa$
parameter as a  measure of the probability that  a monolingual speaker
becomes bilingual, since  $\kappa$ is only an indicator  of the degree
of similarity among  languages, and neglects the role  of their social
status.  It is worth noting  that many bilingual scenarios involve two
highly  differentiated  languages, such  as  Basque  and Castilian  in
northern Spain or Amazigh and Arabic in northern Africa.

How likely  is the  bilingual scenario to  be relevant in  the future?
Recent model  approaches suggest that maintaining  a bilingual society
necessarily requires the maintenance  of status as a control parameter
(Chapel et al.  2010). On  the one hand, preserving language diversity
in a globalized world will  need active efforts when small populations
of speakers  are involved.  But on  the other hand, we  must also take
into  account current  demographic trends  (Graddol, 2004)  which will
need  to  be  incorporated  into  future models  of  language  change.
Against early  predictions suggesting the dominant role  of English as
an  exclusive  language,  the  future looks  multilingual.   Different
languages are  gaining relevance as  their social and  economic status
improves. Moreover,  other interesting tendencies start  to develop as
some languages  (such as English,  Portuguese or Dutch)  spread beyond
their original  geographic domains.  They not  only become mutualistic
(as a bilingual speaker acquires  a higher social status) but can also
develop internal  differentiation.  We should expect in  the future to
see the  emergence of (perhaps uintelligible) dialects  of English, as
it happened with Latin.

\section{Spatial dynamics}

The  exclusion point  resulting from  the Lotka-Volterra  equation and
related models  (such as Abrams-Strogatz's model)  implies that strong
competition  leads  to  diversity  reduction. Within  the  context  of
population dynamics, such result was challenged under the introduction
of spatial  degrees of freedom (Sol\'e  et al., 1993;  see also Sol\'e
and  Bascompte,  2007 for  a  review  of  results).  Spatial  dynamics
involves two  basic components. One  is the reaction  term, describing
how populations interact (for example the previous equations described
above).  The second describes  how populations move through space.  It
is  well  known  that  space  is  responsible  for  the  emergence  of
qualitative changes in dynamical patterns (Turing, 1952; Bascompte and
Sol\'e,  2000; Dieckmann  et  al., 2000).   Competition under  spatial
structure generates a completely novel result: since exclusion depends
on initial  conditions, the two potential attractors  can be (locally)
possible.  Starting from  random initial conditions, different species
or languages can exclude each other at different locations.

The extension of  the Abrams-Strogatz model to space  was performed by
Patriarca  and   Lepp\"anen  (2004)  who   used  a  reaction-diffusion
framework.  The  model considers the local dynamics  of the normalized
densities of speakers  using a given language at  each point ${\bf r}$
in space.  If $\phi_x({\bf r},t)$ and $\phi_y({\bf r},t)$ indicate the
local densities of $x$ and $y$ at a given point and time, they read:
\begin{eqnarray}
{\partial \phi_x({\bf  r},t) \over \partial  t} = F(\phi_x,\phi_y)+D_x
\nabla^2\phi_x({\bf r},t)\\ {\partial \phi_y({\bf r},t) \over \partial
  t} = -F(\phi_x,\phi_y)+D_y \nabla^2\phi_y({\bf r},t),
\end{eqnarray}
where  $F(\phi_x,\phi_y)$  is  just  the  AS equation  for  the  local
densities:
\begin{equation}
F(\phi_x,\phi_y)=s_x    \phi_y    \phi_x^{\alpha}    -   s_y    \phi_x\phi_y^{\alpha},
\end{equation}
and $s_x, s_y$  indicate the status of each  language.  The $D_i$'s on
the  right  side   of  the  equation  are  the   so  called  diffusion
coefficients associated to the spreading process.
\begin{figure}
\centering \scalebox{0.65} {\includegraphics{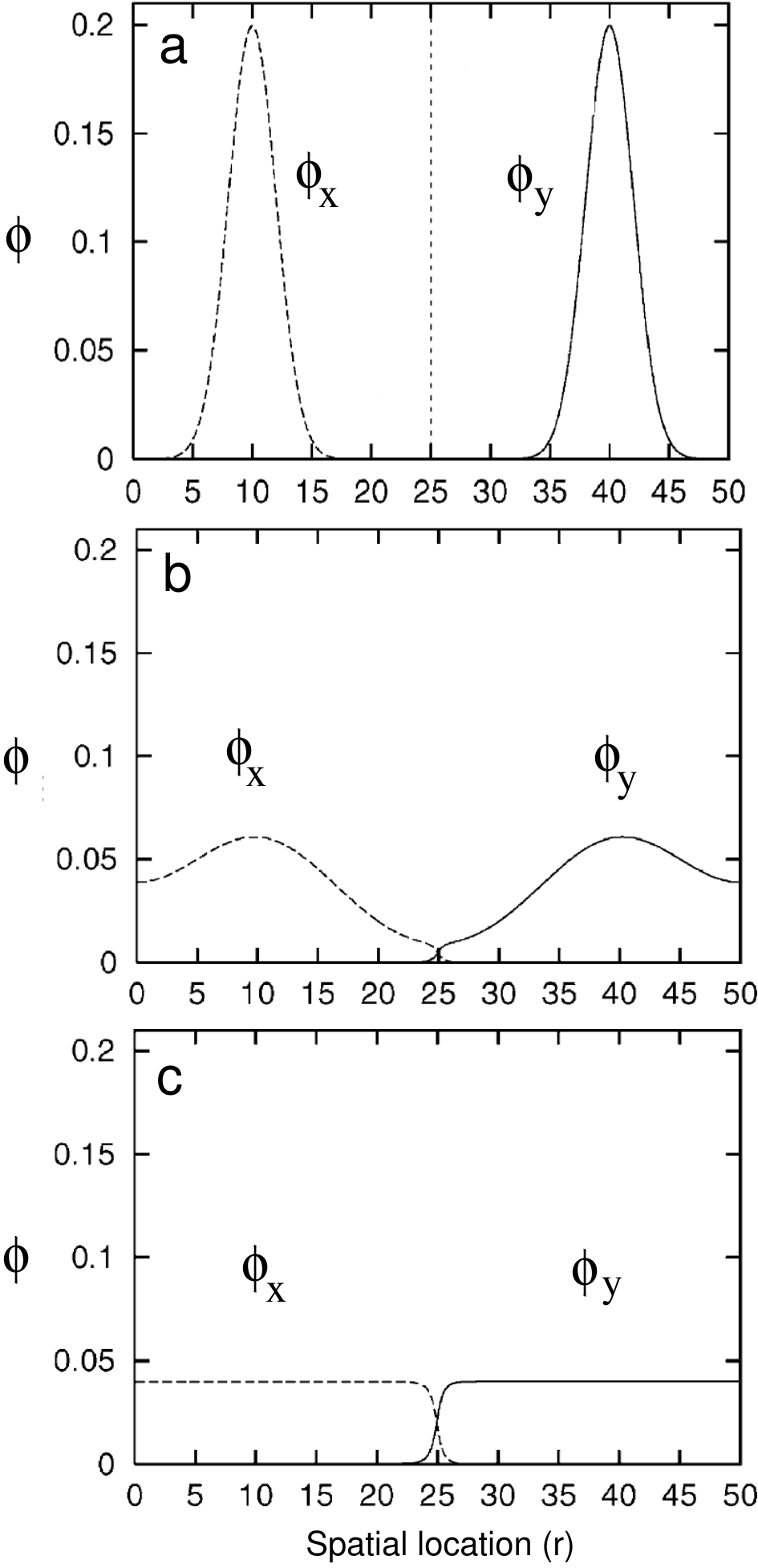}}
\caption{Spatial segregation  of languages over time. Here  we use the
  discretized  equations  of  two  competing  languages  in  order  to
  calculate  their population  of speakers  (relative  frequency) over
  time. We start  in (a) from two segregated  populations of speakers,
  each  in  a different  domain  and  having  a Gaussian  shape,  with
  $N_x(0)=N_y(0)=1/2$,  $\alpha=1.3$ and  status  parameters fixed  to
  $s_x=1-s_y=0.55$.  As  we can see (see text)  although locally there
  is exclusion  of one language,  globally both languages  coexist. As
  time  proceeds  (b-c)  the   spatial  distribution  converges  to  a
  homogeneous state where each  language survives in each domain. Here
  $t(b)=10^3$ and $t_c=10^4$.}
\label{spatial}
\end{figure}
The  previous equations  can be  numerically integrated  (Dieckmann et
al.,  2000).   We will  illustrate  this  by  using a  one-dimensional
spatial   system   (the    generalization   to   two   dimensions   is
straightforward).  First, we  discretize $\partial \phi/\partial t$ as
follows:
\begin{eqnarray}
{\partial \phi_x(r,t) \over \partial t} \approx {\phi(r, t+\Delta t) -
  \phi(r, t) \over \Delta t},
\end{eqnarray}
where  $r$  is  the  local  position  in  the  one-dimensional  domain
$Z=[0,L]$ and  $\Delta t$ some characteristic  time scale.  Similarly,
the discretization of the diffusion term is made as follows:
\begin{eqnarray}
{\partial^2  \phi_x(r,t) \over  \partial  r^2} \approx  {\phi(r+\Delta
  r,t)+\phi(r-\Delta r,t)-2\phi(r, t) \over \Delta r^2},
\end{eqnarray}
being $\Delta r$ the corresponding characteristic spatial scale. Using
these definitions,  we obtain  an equation for  the time  evolution of
$\phi_x(r, t)$:
\begin{eqnarray}
\phi_x(r,   t+\Delta   t)&=&\phi_x(r,   t)  +\nonumber\\   
&&\left [  F(\phi_x,\phi_y)+{D       \over       \Delta      r^2}(\phi(r+\Delta r,t)+\right.\nonumber\\  
&&\left.+\phi(r-\Delta  r,t)-2\phi(r, t))  \right] \Delta t.\nonumber
\end{eqnarray}

Additionally, boundary  conditions need  to be included.   These allow
defining  the impact  of  finite  size effects  and  geography on  the
dynamics and  equilibrium states. The reasonable assumption  is to use
zero-flux (von Neumann) boundary conditions, namely
\begin{equation}
\left ( {\partial \phi_x(r,t) \over \partial r} \right )_{r=0} = \left
( {\partial \phi_x(r,t) \over \partial r} \right )_{r=L} = 0.
\end{equation}
In     terms    of     our    discretization,     we     would    have
$\phi_x(0,t)-\phi_x(\Delta  r,t)=0$  and  $\phi_x(L,t)-\phi_x(L-\Delta
r,t)=0$.

The dynamics  starts with two  populations of speakers located  in two
different domains $Z_x$ and $Z_y$  (so that $Z=Z_y \cup Z_y$). This is
shown in  figure 4a,  where we display  the initial condition.   If we
label as $N_x^{\mu}$ and $N_y^{\mu}$ the total populations of speakers
in each domain $\mu=1,2$, at a given domain $Z_{\mu}$ we would have:
\begin{equation}
N_i^{\mu} (t) = \int_{Z_{\mu}} \phi_i(r,t)dr,
\end{equation}
starting from $N_i=1/2$ following  a Gaussian shape (see Patriarca and
Lepp\"anen, 2004). As the dynamics proceeds, we can observe a tendency
towards maintaining the  spatial seggregation.  Each language ``wins''
in its initial domain, and  eventually both reach a homogeneous steady
state  within such  domain. Generalizations  to  heterogeneous domains
reveal that the  previous patterns can be affected  by both historical
events  and spatial  inhomogeneities (Patriarca  and  Heinsalu, 2008).
However, the main message from  this approach is robust and completely
related to models of competing  populations in ecology (Sol\'e et al.,
1993; Sol\'e and  Bascompte 2006). In summary, this  tells us that the
effects  of spatial  degrees of  freedom on  language dynamics  have a
great impact on the coexistence {\em versus} extinction scenarios.

Space slows down the  effects of competitive interactions, effectively
reducing  competition  at the  local  scale.   Moreover,  the role  of
diffusion  (dispersal)  on   competition  dynamics  allows  to  create
well-defined domains  where given  languages or species  have replaced
others. In this context, it  is clear that the increasing connectivity
of  our world  due  to globalization  has  made easier  to reduce  the
potential  impact of  geography  in the  propagation  of languages  or
epidemics (Buchanan,  2003). Although we do live  in a two-dimensional
surface, the world has  certainly changed and spatial constraints have
been strongly reduced.

\section{String models of language change}

As already  mentioned in section 2,  a collection of  words provides a
first definition of a language in terms of its lexicon. This of course
ignores a crucial component  of language: words interact in non-random
ways  and higher-order  levels of  organization should  be  taken into
account. However, as it occurs with some theoretical models of diverse
ecosystems (Sol\'e and Bascompte, 2006) some relevant problems such as
diversity and  its maintenance can  be properly addressed  by ignoring
interactions. Following this picture,  we consider in this section the
lexical component of  language viewed as a bag of words  and how a set
of languages competing  for a given population of  speakers can evolve
towards  a  single,  dominant  tongue  or instead  a  diverse  set  of
coexisting languages.
\begin{figure}
\centering \scalebox{0.4} {\includegraphics{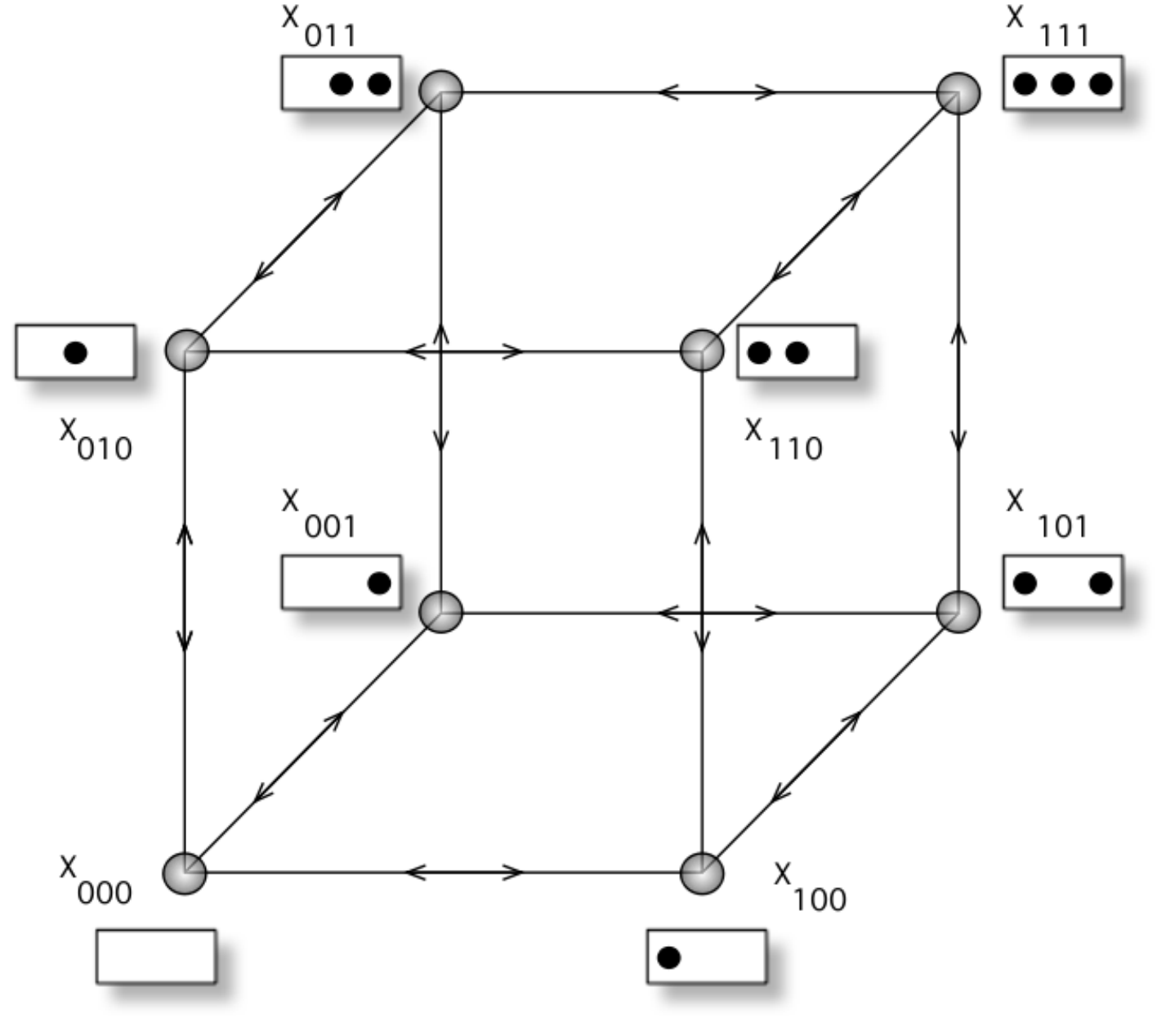}}
\caption{String language model. Here a given set of elements defines a
  language. Each (possible)  language is defined by a  string of $\nu$
  bits (here $L=3$)  and thus $2^L$ possible languages  are present in
  the hypercube.   The two types  of elements are indicated  as filled
  ($1$) and empty ($0$) circles, respectively.}
\label{langcube}
\end{figure}

A  fruitful toy model  of language  change is  provided by  the string
approximation  (Stauffer  et  al.,  2006;  Zanette,  2008).   In  this
approach, each  language ${\cal L}_i$  is treated as a  binary string,
i. e.   ${\cal L}_i=(S^i_1, S^i_2,  ..., S^i_L)$ of length  $L$.  Here
$S^i_j \in  \{0,1\}$ and, as defined,  a finite but very  large set of
potential  languages exists.  Specifically, a  set of  languages ${\bf
  \cal L}$ is defined, namely
\begin{equation}
{\bf \cal L} = \{ {\cal L}_1, {\cal L}_2, ..., {\cal L}_M \},
\end{equation}
with  $M=2^L$. These languages  can be  located as  the vertices  of a
hypercube,  as shown  in figure  5 for  $L=3$.  Nodes  (languages) are
linked  through  arrows  (in  both  directions)  indicating  that  two
connected languages differ in a single bit. This is a very small sized
system.   As $L$  increases,  a combinatorial  explosion of  potential
strings takes place.

\subsection{Mean field model}

A given language  ${\cal L}_i$ is shared by  a population of speakers,
to  be indicated  as  $x_i$, and  such  that the  total population  of
speakers using any  language is normalized (i. e.   $\sum_i x_i=1$). A
mean field model  for this class of description  has been presented by
Damian  Zanette,   using  a  number  of   simplifications  that  allow
understanding  the  qualitative  behavior  of competing  and  mutating
languages (Zanette, 2008).  A few  basic assumptions are made in order
to construct the model. First,  a simple fitness function $\phi(x)$ is
defined.   This  function  measures  the likelihood  of  abandoning  a
language.   This  is a  decreasing  function  of  $x$, and  such  that
$\phi(0)=1$ and $\phi(1)=0$. Different choices are possible, including
for example  $1-x, 1-x^2$ or  $(1-x)^2$. On the other  hand, mutations
are also included:  a given language can change  if individuals modify
some of their bits.

The  mean field  model  considers the  time  evolution of  populations
assuming no  spatial interactions. If we indicate  ${\bf x}=(x_1, ...,
x_M)$,  the  basic  equations  will  be  described  in  terms  of  two
components:
\begin{equation}
  {dx_i \over dt} = A_i({\bf x}) - M_i({\bf x}),
\end{equation}
where both language abandonment  $A_i({\bf x})$ and mutation $M_i({\bf
  x})$ are introduced.  Specifically, the following choices are made:
\begin{equation}
A_i({\bf  x}) =  \rho x_i  \left (  \langle \phi  \rangle  - \phi(x_i)\right ),
\end{equation}
for the population  dynamics of change due to  abandonment.  This is a
replicator  equation, where  the speed  of  growth is  defined by  the
difference between average fitness $ \langle \phi \rangle$, namely
\begin{equation}
 \langle \phi \rangle = \sum_{j=1}^N \phi(x_j) x_j,
\end{equation}
and the actual fitness $\phi(x_i)$ of the $i$-language. Here $\rho$ is
the recruitment rate (assumed to be equal in all languages). What this
fitness  function  introduces is  a  multiplicative  effect: the  more
speakers that  use a  given language, the  more likely that  they keep
using  it and  others join  the same  group.  Conversely,  if  a given
language is rare, its speakers  might easily shift to some other, more
common language.
\begin{figure*}
\centering \scalebox{0.6} {\includegraphics{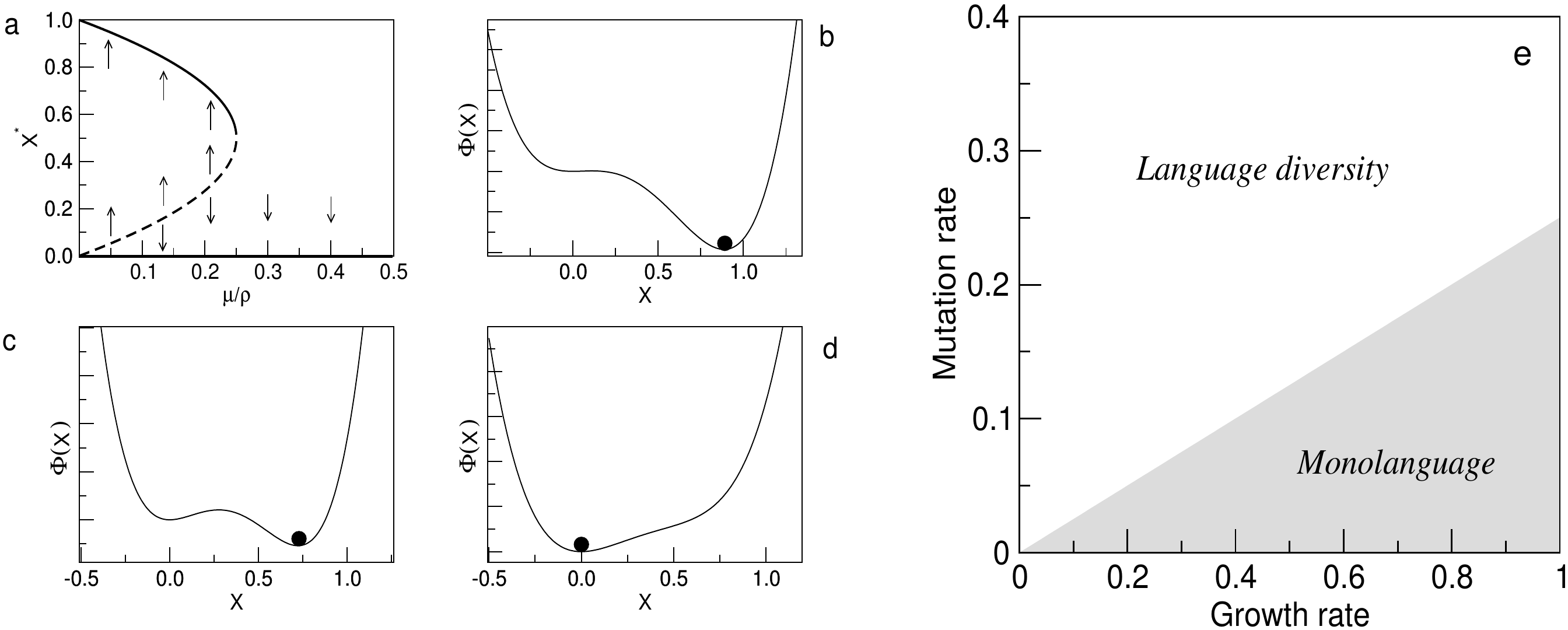}}
\caption{Phase  transitions as  bifurcations in  Zanette's  mean field
  model of  supersymmetric language competition.   In (a) we  show the
  bifurcation diagram using $\mu/\rho$  as control parameter.  Once we
  cross   the  critical   point,  a   sharp  transition   occurs  from
  monolanguage  to   language  diversity.   This   transition  can  be
  visualized  using  the  potential  function $\Phi(x)$  whose  minima
  correspond to possible equilibrium points. Here we use $\rho=1$ with
  (b) $\mu=0.1$, (c) $\mu=0.2$ and  (d) $\mu=0.3$. In (e) we also plot
  the phase diagram using the $(\rho,\mu)$ parameter space.}
\label{ZMFT}
\end{figure*}

The second  term includes  all possible flows  between ``neighboring''
languages.  It is defined as:
\begin{equation}
M_i({\bf x})  = {\mu \over  L} \left (  \sum_{j=1}^N W_{ij} x_j  - x_i
\sum_{j=1}^N W_{ji} \right ).
\end{equation}
In this  sum, we introduce  the transition rates $W_{ij}$  of mutating
from  language  ${\cal  L}_i$   to  language  ${\cal  L}_j$  and  vice
versa. Only single  mutations are allowed, and thus  $W_{ij}=1$ if the
Hamming  distance $D({\cal  L}_i,{\cal  L}_j)$ is  exactly $1$.   More
precisely, if
\begin{equation}
D({\cal L}_i,{\cal L}_j) = \sum_{k=1}^L \vert S^i_k - S^j_k \vert = 1.
\end{equation}
In other words, only  nearest-neighbor movements through the hypercube
are  allowed.  In  summary,  $A({\bf x})$  provides  a description  of
competitive interactions  whereas $M({\bf x})$  gives the contribution
of  small   changes  in   the  string  composition.    The  background
``mutation''  rate  $\mu$  is  weighted  by  the  matrix  coefficients
$W_{ij}$  associated with the  likelihood of  each specific  change to
occur.

This model is a general description of the bit string approximation to
language dynamics.  However, the  general solution cannot be found and
we need to  analyse simpler cases. An example is  provided in the next
section.  Although the assumptions are rather strong, numerical models
with  more  relaxed assumptions  seem  to  confirm  the basic  results
reported below.

\subsection{Supersymmetric scenario}

A  solvable  limit  case  with  obvious  interest  to  our  discussion
considers a  population where a  single language has a  population $x$
whereas   all  others   have   a  small,   identical   size,  i.    e.
$x_i=(1-x)/(N-1)$.    The  main  objective   of  defining   such  {\em
  supersymmetric}  model is  making the  previous system  of equations
collapse  into  a single  differential  equation,  which  we can  then
analyze. In particular, we want to determine when the $x=0$ state will
be observed, meaning that no single dominant language is stable.

Since we have the normalization condition, now defined by:
\begin{equation}
\sum_{j=1}^N x_j =  x + \sum_{j=1}^{M-1} \left  ( {1 - x \over  N - 1}
\right ) = 1
\end{equation}
(where  we choose $x$  to be  the $M$-th  population, without  loss of
generality).  In this case the average fitness reads:
\begin{eqnarray}
 \langle \phi \rangle = \phi(x) x + \sum_{j=1}^{M-1} \phi \left ( {1 -
   x \over N - 1} \right )\left ( {1 - x \over N - 1} \right ).
\end{eqnarray}
Using the special linear case $\phi(x)=1-x$, we obtain:
\begin{eqnarray}
A(x) = \rho x (1-x) \left ( x - {1 - x \over N - 1} \right ).
\end{eqnarray}
The  second term  is  easy to  obtain:  since $x$  has  (as any  other
language) exactly $L$ nearest neighbors, and given the symmetry of our
system, we have:
\begin{eqnarray}
B(x) = {\mu \over L} \left ( L {1 -  x \over N - 1 } - xL \right ) = -
\mu \left ( {Nx - 1\over N - 1 } \right ).
\end{eqnarray}
And the final equation for $x$ is thus, for the large-$N$ limit (i. e.
when $N \gg 1$):
\begin{equation}
{d x \over dt} = \rho x^2 (1-x) - \mu x.
\end{equation}
This equation  describes an interesting  scenario where growth  is not
logistic, as it happened with  our previous model of word propagation.
As  we can  see, the  first  term in  the right-hand  side involves  a
quadratic component,  indicating a self-reinforcing  phenomenon.  This
type   of  model   is  typical   of  systems   exhibiting  cooperative
interactions  and  an   important  characteristic  is  its  hyperbolic
dynamics:  instead   of  an  exponential-like   approximation  to  the
equilibrium state, a very fast approach takes place.

The  model has  three equilibrium  points: (a)  the  extinction state,
$x^*=0$  where the  large language  disappears; (b)  two  fixed points
$x_{\pm}^*$ defined as:
\begin{equation}
x_{\pm}^* = {1  \over 2} \left (  1 \pm \sqrt{1 - {4  \mu \over \rho}}
\right ).
\end{equation}
As we can see, these two fixed points exist provided that $\mu < \mu_c
=  \rho/4$.   Since three  fixed  points  coexist  in this  domain  of
parameter space,  and the trivial  one ($x^*=0$) is stable,  the other
two points, namely  $x_-^*$ and $x_+^*$, must be  unstable and stable,
respectively.  If $\mu<\mu_c$,  the upper branch $x_+^*$ corresponding
to a monolingual solution, is stable.
 
In figure 3a  we illustrate these results by  means of the bifurcation
diagram using $\rho=1$ and different  values of $\mu$. In terms of the
potential function we have:
\begin{equation}
{d x \over dt} = - {\partial \Phi_{\mu}(x) \over \partial x },
\end{equation}
where  $\Phi_{\mu}(x)  =-\int  (A(x)-B(x))dx$,  which for  our  system
reads:
\begin{equation}
\Phi_{\mu}(x)= - \rho \left ( {x^3 \over 3} + {x^4 \over 4} \right ) +
\mu {x^2 \over 2}.
\end{equation}
In fig 5a-d  three examples of this potential are  shown, where we can
see that  the location  of the equilibrium  point is shifted  from the
monolanguage  state to  the  diverse  state as  $\mu$  is tuned.   The
corresponding phases in the  $(\rho,\mu)$ parameter space are shown in
figure 5.

It is interesting  to see that this model and  its phase transition is
somewhat connected  to the error  threshold problem associated  to the
dynamics of RNA viruses (Domingo et al. 1995; Eigen et al., 1987). For
a  single language  to  maintain  its dominant  position,  it must  be
efficient in  recruiting and  keeping speakers. But  it also  needs to
keep heterogeneity (resulting from  ``mutations'') at a reasonable low
level.  If  changes go  beyond a given  threshold, there is  a runaway
effect that eventually pushes the  system into a variety of coexisting
sub-languages.  An error  threshold is thus at work,  but in this case
the transition  is of first  order.  This result would  indicate that,
provided that a  source of change is active  and beyond threshold, the
emergence of multiple uninteligible tongues should be expected.

String  models   of  this  type   only  capture  one  layer   of  word
complexity. Perhaps  future models  will consider ways  of introducing
further  internal layers of  organization described  in terms  of {\em
  superstrings}. Such  superstring models should be  able to introduce
semantics,  phonology and  other key  features  that are  known to  be
relevant. An  example in this direction  is provided by  models of the
emergence of linguistic categories (Puglisi et al., 2008).

\section{Global patterns and scaling laws}

Tracking the relative importance  of languages and in particular their
likelihood of getting extinct requires having the appropriate censuses
of number  of speakers using  each language. The  statistical patterns
displayed  by languages  in their  spatial and  demographic dimensions
provide further  clues for the  presence of non-trivial  links between
language and ecology (Nettle 1998;  Pagel and Mace, 2004; Pagel 2009).
These patterns  also provide a  large-scale picture of  languages, not
restricted  to  small  geographical  domains or  countries.   In  this
section we consider two of  such statistical patterns. It is important
to  notice  that,  strictly   speaking,  this  problem  involves  both
ecological  and evolutionary time  scales. In  a given  ecosystem, the
succession  process leading  to  a mature,  diverse  community can  be
described in terms of ecological dynamics. At this level, invasion and
network   species  interactions  are   both  relevant.   However,  the
composition  of  the   local  pool  of  species  is   the  outcome  of
evolutionary dynamics.

Some spatial models of language change have been presented in order to
explain  the results  shown below  (see de  Oliveira et  al.  2005; de
Oliveira  et  al,  2008).    The  close  correlation  between  species
diversity  and language  richness,  as reported  by different  studies
(Mace and Pagel, 1995; Moore  at al., 2002; Gaston 2005) suggests that
some rules  of organization  might be common.  As an example,  a large
scale study of correlations  among biological species and cultural and
linguistic diversity in Africa (Moore  et al., 2002) revealed that one
third  of  language  richness  can   be  explained  on  the  basis  of
environmental factors. These included rainfall and productivity, which
were  shown   to  affect  the   distributions  of  both   species  and
languages. However, there are  also important differences that need an
explanation.

\subsection{Species-area relations}

One of the universal laws  of ecological organization is the so called
species-area  relation (Rosenzweig,  1995).  It  establishes  that the
diversity $D$ (measured as the number of different species) in a given
area $A$ follows a power law
\begin{equation}
D \sim A^z,
\end{equation}
where  the exponent  $z$ tipically  varies from  $z=0.1$  to $z=0.45$.
Interestingly, languages seem to  follow similar trends.  They exhibit
an enormous diversity, strongly  tied to geographical constraints.  As
it occurs  with species  distributions, languages and  their evolution
are shaped by the presence  of physical barriers, population sizes and
contingencies of  many kinds.  In  this context, differences  are also
clear: speciation in ecosystems can take place without the presence of
physical  barriers, whereas  some type  of population  isolation seems
necessary for one language to  yield two diferent languages, i.e., two
linguistic  variants that  are  not fully  interintelligible.  On  the
other hand, there is a  continuous drift in both species and languages
that  makes  them  change.   A  second  difference  involves  the  way
extinction occurs.  Species  get extinct once the last  of its members
is gone.   Languages get extinct too  once they are  not used anymore,
even if its native speakers are still alive (Dalby, 2005).

\begin{figure}
\centering \scalebox{0.45} {\includegraphics{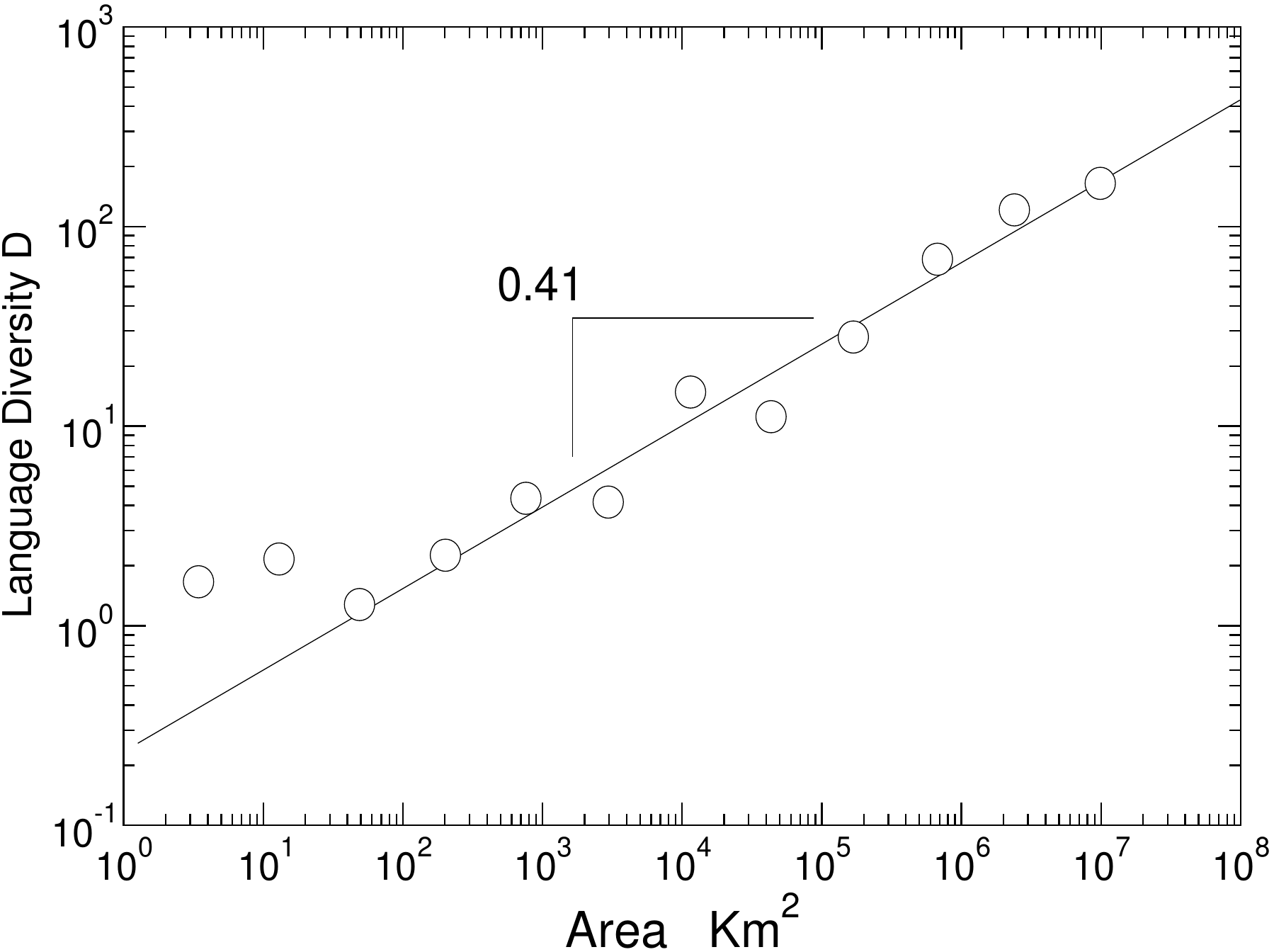}}
\caption{Scaling law in the  distribution of language diversity $D$ as
  a function of  area. The best fit  to the power law $D  \sim A^z$ is
  shown. Redrawn from Gomes et al., 1999.}
\label{scaling1}
\end{figure}

Studies  of  geographical  patterns  of language  distribution  reveal
complex phenomena at multiple scales. As an example, it was shown that
they also  display a  diversity-area scaling law,  with $z =  0.41 \pm
0.03$ (Gomes et  al., 1999).  In figure 7 we show  the results of this
analysis for a compilation listing  more than 6700 languages spoken in
228 countries.  The  power law fit is very good  and spans over almost
six decades (with a deviation for areas smaller than $30 Km^2$) (Gomes
et al., 1999).  Similar results  are obtained by using population size
$N$ instead  of areas. In this case,  it was shown that  the new power
law reads:
\begin{equation}
D \sim N^{\nu}
\end{equation}
with  $\nu =  0.50 \pm  0.04$.  However,  a close  inspection  of data
reveals the  impact of other  forces acting on language  diversity. An
example is  the contrast between  Europe and New Guinea  (see Diamond,
1997 and references therein).  The former has $10^7 Km^2$ and includes
63  languages, whereas the  later, with  only less  than one  tenth of
Europe's  surface, contains  around $10^3$  different  languages.  The
singularity  of  New  Guinea  has  been  carefully  analysed  by  many
authors. Take for example Papua  New Guinea, which contains just $0.1$
percent  of the  world's  population  but more  than  $13$ percent  of
world's  languages.   It  is  geographically  an  extremely  irregular
landscape,  which   creates  multiple  opportunities   for  isolation.
Moreover,    $80$    percent   of    its    land    is   covered    by
rainforests. Additionally, food production is continuous, with no food
shortages  and a good  yield.  Bilingualism  is widespread,  with most
speakers of the  dominant Tok Pisin also speaking  some local language
too (being exposed to several).  Given the high yields of food harvest
together  with  biogeographical  constraints,  there has  been  little
incentive to create large-scale  trade. A consequence of such scenario
is a dynamic equilibrium far from language homogeneization (see Nettle
and Romaine, 2000 for a review).

The  species-area relation  has been  explained  in a  number of  ways
through   models    of   population   dynamics    on   two-dimensional
domains. Beyond their differences,  these models share the presence of
stochastic dynamics  involving multiplicative processes.   In ecology,
such  type of  processes are  characterized by  positive  and negative
demographical  responses  proportional   to  the  current  populations
involved: a  larger population  will be more  likely to  increase, but
also more  likely to suffer the  attack of a given  parasite (and thus
experience  a rapid  decline). Within  language,  the rich-gets-richer
effect is  obvious, whereas  there is no  equivalent for  the negative
effects of ``parasitic'' languages.

\subsection{Language richness laws}

A  different  measure  of  language diversity  involves  the  language
richness among different countries.  If ${\cal N}(D)$ is the frequency
of  countries  with $D$  diferent  languages  each,  we can  plot  the
cumulative distribution ${\cal N}_>(D)$ defined as:
\begin{equation}
{\cal N}_>(D) = \int_D^{\infty} {\cal N}(D) dD.
\end{equation}
The resulting  plot is rather  interesting (fig 8a):  the distribution
follows a two-regime scaling behavior, i. e.
\begin{equation}
{\cal N}_>(D) \sim D^{-\beta},
\end{equation}
with $\beta=0.6$ for $6<D<60$  and $\beta=1.1$ for $60<D<700$. What is
revealed from this plot?  The first domain has an associated power law
with  a  small  exponent  (here  ${\cal N}(D)  \sim  D^{-1.6}$):  many
countries have a  small language diversity. But once  we cross a given
threshold  $D  \approx 60$  the  decay  becomes  faster. One  possible
interpretation is  that countries having  a very large  diversity will
have  harder   times  to  preserve   their  unity  under   the  social
differentiation  associated   to  ethnic  diversity   (Gomes  et  al.,
1999). 
\begin{figure*}
\centering \scalebox{0.6} {\includegraphics{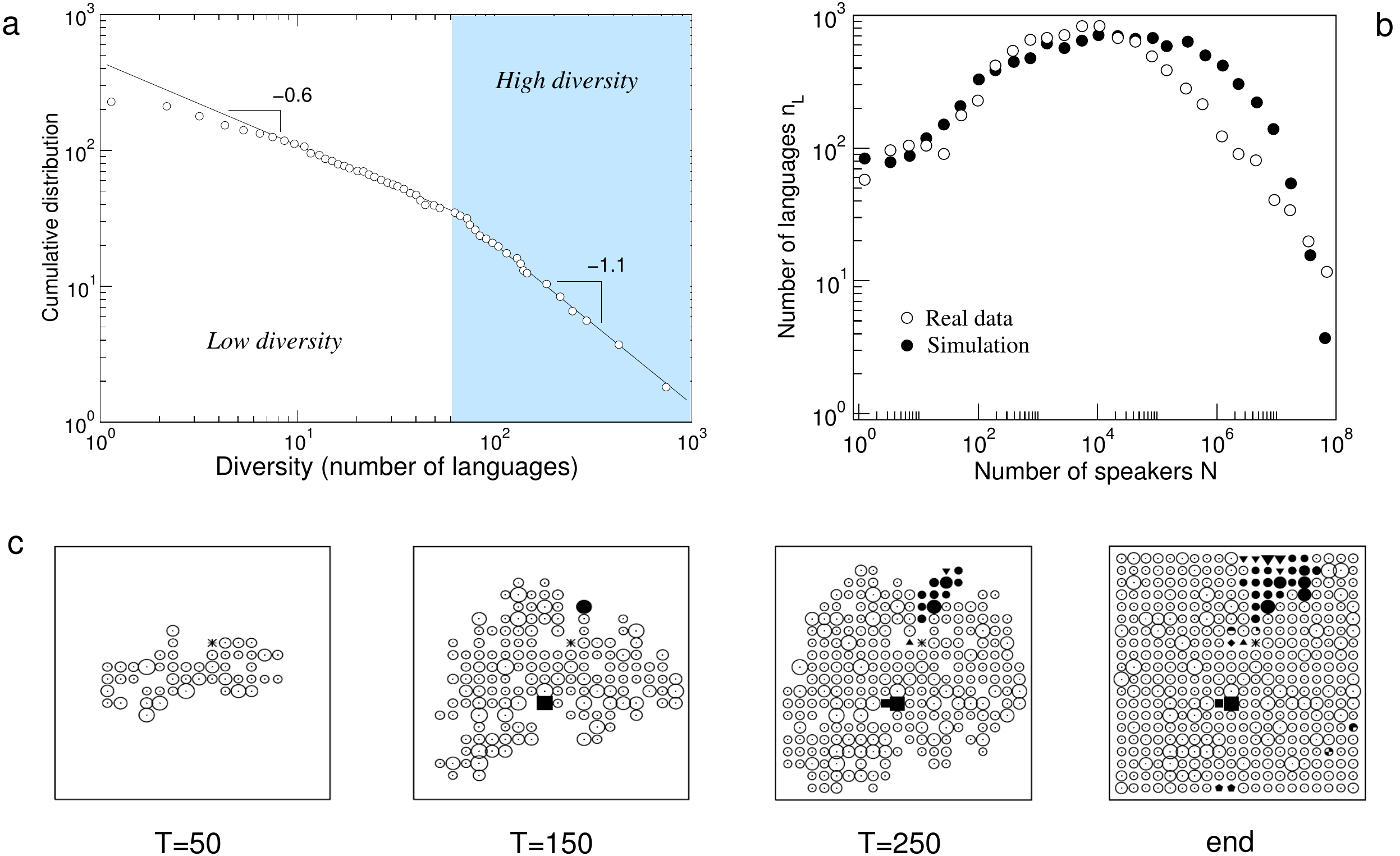}}
\caption{Scaling  laws in language  diversity.  (a)  Here we  plot the
  cumulative distribution  of languages using the  number of countries
  with a language  diversity greater than $D$.  Redrawn  from Gomes et
  al.,  1999. The marked  area indicates  the domain  of language-rich
  countries,  whose  distribution is  steeper  than the  low-diversity
  domain. (b) Distribution of  languages having $N$ speakers. Here the
  data  set  for languages  is  compared  with  a simulation  using  a
  specific set of  parameters (see de Oliveira et  al, 2008). Although
  different  parameter  sets give  different  curves, the  qualitative
  behavior is  allways the same.  In  (c) we show four  snapshots of a
  model of  language diversity  dynamics on a  two-dimensional lattice
  (adapted  from de  Oliveira et  al.,  2008). Here  each symbol  type
  indicates one  given language, whereas its size  indicates the local
  population  allowed.  As  time  proceeds, mutations  arise  and  new
  languages emerge and spread (see text).}
\end{figure*}

A related  distribution is given  by the number of  languages $n_L(N)$
with a  population size  of $N$  speakers. In figure  8b we  display a
log-log plot of the data  set (after binning) which shows a log-normal
behavior,  with  an enhanced  number  of  small-sized languages.  This
pattern (as well  as the scaling with area) is  reproduced by a simple
model presented below.

\subsection{Language diversity model}

A simple spatial model has been proposed in (de Oliveira et al., 2008)
as an extension  of previous work (de Oliveira et  al., 2006; see also
Silva and  Oliveira, 2008). The  model combines a  stochastic cellular
automaton   approach   with   non-local   rules   and   a   bit-string
implementation. Starting  from an empty lattice $\Omega$  of $L \times
L$ sites.  Each site $(i,j) \in  \Omega$ is characterized  by a random
number $1  \le K_{ij} \le M$ (with  uniform distribution) representing
the  maximum  population  of   speakers  achievable  by  the  language
occupying it  (the carrying capacity). Only one  language ${\cal L}_i$
can be present at a given site and (as in section 6) is represented by
a string  ${\cal L}_i=(S^i_1,  S^i_2, ..., S^i_L)$  of length  $L$.  A
seed ${\cal  L}_1$ is located at  $t=0$ at a given  site $(a,b)$, thus
having a  population $K_{ab}$. Now  dispersal to nearest  neighbors in
the lattice  occurs, favouring the spread towards  sites having higher
$K_{ij}$. Moreover,  at a given  site the given language  ${\cal L}_k$
can  change  (mutate)  to  a  new  one with  a  probability  $\mu_k  =
\alpha/f({\cal L}_k)$. Here $f({\cal  L}_k)$ is the fitness associated
to ${\cal L}_k$, here chosen as:
\begin{equation}
f({\cal L}_k) = \sum_{i,j} K_{ij} \theta({\cal L}(i,j), {\cal L}_k)
\end{equation}
with  $\theta(m,n)=1$  if $m=n$  and  zero  otherwise.  In words,  the
fitness considers  the total  occupation of the  lattice (in  terms of
speakers),  and  the  likelihood  of  a language  to  mutate  is  thus
size-dependent following  an inverse law.  In this way  we incorporate
the  well known  fact  that  the impact  of  mutations favour  genetic
drift. The previous  rules allow a diverse set  of languages to expand
and  eventually occupy  the whole  lattice.   An example  is shown  in
figure  8b for  a small  ($L=50$) lattice.  We can  see  how languages
emerge and spread around, generating monolingual patches.

In spite of  its simplicity and strong assumptions,  the model is able
to  capture  several  qualitative   properties  of  both  spatial  and
statistical power laws, similar  to those presented above (de Oliveira
et al., 2006; 2008).  In some sense, we can conclude that the observed
commonalities  point  towards  shared system-level  properties.   This
conclusion is partially true: the process of ecosystem building can be
understood   in  terms   of  a   spatial  colonization   of  available
patches. Each  patch offers a given  range of conditions  that make it
more or less  suitable for the colonizer to  persist.  If colonization
occurs  locally,   nearest  patches  will  be   occupied  by  best-fit
competitors\footnote{In fact  two opposite strategies  can be observed
  in nature, particularly when  looking at the colonization of habitat
  by plants, which can invest either in a few, well-protected seeds or
  many, small ones. In the second case, most of the seeds will fail to
  survive.}.   In  an  ecological-like model,  non-local  colonization
events will occur due to the introduction of species from the regional
pool (see Sol\'e et al, 2002) but these events can also be interpreted
as speciations.   Perhaps the most obvious  difference with ecological
models is  the assumption of a  fitness trait that  involves the whole
population of the species. Such a non-local effect seems reasonable to
assume   when  thinking  of   language  as   a  vehicle   of  economic
influence. Larger  communties of speakers  are likely to be  much more
efficient in further expanding.

\section{Discussion}

Language dynamics has attracted  the attention of physicists, computer
scientists and  theoretical biologists alike as  a challenging problem
of complexity (Gomes et al., 1999; Smith, 2002; Steels, 2005; Stauffer
and Schulze 2005; Brighton et  al. 2005; Baxter et al., 2006; Kosmidis
et al.,  2006; Lieberman et al.,  2007; Schulze et  al., 2008; Zanette
2008; Cattuto  et al.,  2009; Gong  et al 2008).  Language makes  us a
cooperative  species   and  has  been  crucial   to  our  evolutionary
success. It pervades all aspects  of human society.  Its complexity is
extraordinary  and it  would be  easy to  conclude that  any modelling
effort  will end in  failure. However,  as it  occurs with  many other
complex systems, important features of language structure and dynamics
can be  captured by means of simple  models. The fact that  we live in
the midst  of a rapid  globalization process makes the  development of
such models an important task.

In this work we have  explored the application of several methods from
nonlinear  dynamics and  statistical physics  to different  aspects of
language  dynamics.   Many  of  the  above  described  models  can  be
interpreted also in the light of ecological dynamics, generally taking
species instead  of languages. In  this last section we  shall discuss
the scope  of such  an analogy, focusing  our attention on  some basic
similarities and differences between  linguistics and ecology. Some of
these  are  summarized in  table  1.   Some  differences are  obvious.
Species are  embedded within  complex ecosystems defining  networks of
species  interactions  (Montoya  et  al,  2006).  Such  webs  are  the
architecture of  ecological organization. Although one  could define a
matrix  of   language-language  interaction  in   terms  of  dominance
relations of some sort, the equivalence would be weak. Similarly, some
dynamical  processes known  to  play important  roles  in ecology  are
absent in  language dynamics.  A  dramatic example is provided  by the
impact  of small  invasions of  alien  species introduced  in a  given
ecosystem.  Very  often, the invaders  expand rapidly and  trigger the
collapse  of the whole  community.  A  small group  of humans  using a
foreign language would  not succeed to propagate within  a much larger
community of speakers, unless a huge assymetry among the social status
is at work.
\begin{table*}
\begin{center} 
\begin{tabular}{| c | c | c |}
\hline &&\\ 
& Species &  Languages\\ 
&&\\ 
\hline &&\\ 
Nature & Classes of living beings & Community-shared codes \\ 
&& \\ 
Separation based on & Lack  of interbreeding  & Unintelligibility \\  
&& \\  
Origination & Genetic/geographic isolation &  Geographic barriers \\ 
&&\\ 
Extinction causes  &  Competition/external   events  &  Competition/External Events\\ 
&&\\ 
Abundance & Two-regime  scaling & Scaling law \\ 
&&\\ 
Intermediate forms &  Subspecies & Dialects \\ 
&&\\  
Spatial distribution & Species-area law & Language-area scaling \\ 
&&\\ 
Change through time & Gradual+Punctuated  &  Gradual+punctuated  \\  
&&\\ 
Effects  of  small invasion  & Very important  & Rare \\  
&&\\ 
Mutualism  & Very important  & No \\  
&&\\ 
Multilingualism  & No  & Very important \\  
&&\\ 
Network  structure &  Yes  & No\\ 
&&\\ 
\hline
\end{tabular}
\end{center}
\unboldmath
\caption{A comparative list of  features relating the organization and
  change  of languages  and species.   The list  of mechanisms  is not
  exhaustive:   it   only   considers  mainstream   phenomena.    Some
  parallelisms between languages and species should be considered with
  attention. Although small invasions  have deep impact in ecosystem's
  organization,  this factor  rarely  has a  remarkable effect  within
  large  linguistic  communities.  This  is  arguably  related to  the
  tendency that an invading language  displays at the same time, a low
  demographic weight and a low  social status.  It is also interesting
  to observe that mutualism, i.e., a cooperative strategy for survival
  that benefits two or more  species, is completely absent in language
  dynamics. On  the contrary,  multilingualism, as well  as disglossia
  and related phenomena -see text- are features exclusive to language.
  Finally, we empahsize  that analogies to food webs  are difficult to
  define in the study of  language contact. However, it is conceivable
  some  kind of  network abstraction  to represent  the socio-cultural
  relations among languages or communities of speakers.}
\label{Ecotable}
\end{table*}

One  of the  most important  links  between languages  and species  is
strongly  tied to  the  concept  of species  and  its similarity  with
language.   As  is  well-known,  a  group  of  organisms  is  said  to
constitute a species  when they are capable of  interbreeding and they
are separated  from another group  also capable of  interbreeding with
which  they cannot  interbreed.   A  community is  said  to possess  a
language  when   their  members   can  communicate  with   each  other
efficiently using linguistic signs  and they cannot communicate with a
different community  which possesses a different  language.  These two
conceptions  are known  to  be problematic:  there  is, for  instance,
variation  in  the degree  of  success  of  hybridization between  two
species  and  in  the  degree  of  mutual  understanding  between  two
languages.   As  for linguistic  variants,  it  is  not uncommon  that
members  of a  community  A  understand the  linguistic  variant of  a
community B  better than  the members of  B understand  the linguistic
variant of A,  and quite often the decision  of whether two linguistic
variants  constitute a  language or  a dialect  is not  guided  by the
interintelligibility  criterion but  by political  reasons. Therefore,
the boundaries among groups of organisms and among linguistic variants
as  to   interbreeding  and  interintelligibility   are  fuzzy.   Both
languages and species constitute continua where the relative degree of
interintelligibility and interbreeding vary substantially depending on
how close two languages or species are in the continuum.

Competition is  also a crucial  concept to understand  both ecological
and  language dynamics.  Whereas  species in  contact may  compete for
limited resources, languages in contact  may compete for the number of
speakers.   Since languages  are not  constituted of  individuals, but
they are abstract  systems (codes) shared by a  community, it may seem
that  languages  compete  for  the   number  of  speakers  only  in  a
metaphorical sense.   However, it  is remarkable that  the competition
among   languages   and  the   competition   among   species  can   be
mathematically modeled  using similar methods.   At this point,  it is
necessary to take  into consideration the importance of  the role of a
given language  as a social status parameter  in language competition,
provided  that  different  languages  may  distribute  differently  in
society,  but  not  different  species  in  an  ecosystem.   Moreover,
competition among  different languages in contact  can be materialized
in many  different ways,  depending on how  a given  culture conceives
mono/multilingualism.

Although the  ecological metaphor of language dynamics  fits well with
several important features, there are a number of important linguistic
phenomena which have no equivalent in ecology.  Some
members of  a community may  be bilingual or multilingual,  i.e., they
may  possess  not  only  the  traditional language  of  the  community
(namely, their  mother tongue), but also other  languages or dialects.
Indeed, some  members of  a community may  use different  languages or
dialects in different social  spheres, a phenomenon called disglossia.
It is also worth noting that, when speakers of multiple languages have
to  communicate and  do  not have  the  chance to  learn each  other's
language, they develop a simplified code, a pidgin, which may increase
its  degree of complexity  over the  years. However,  when a  group of
children  are exposed  to a  pidgin  at the  age when  they acquire  a
language, they  transform it  into a full  complex language,  a creole
(DeGraff, 1999, and references therein). In this context, although some parallels 
have been traced between creolization and genetic hybridization in 
plants (Croft 2000) they don't seem well supported or even properly defined. 

Another  related   and  remarkable  linguistic   idiosyncracy  is  the
emergence  of  new languages  ex  nihilo.  This  is  the  case of  the
Nicaraguan  sign  language (Kegl  et  al.,  1999) which  spontaneously
developed among deaf school children in western Nicaragua over a short
period of  time once deaf individuals (until  then growing essentially
isolated) could  start communicating to  each other.  Starting  from a
very limited number of signs and unable to learn Spanish, it was found
that the  group rapidly  developed a grammar,  which became  a complex
language at  the second ``generation'', as  soon as the  next group of
children  learned it  from  the  first one.  A  similar situation  was
analysed for the Al-Sayyid Bedouin  Sign Language, which has arisen in
the last 70 years within an isolated community (Sandler et al., 2005).
This type of phenomena highlights  the role of the cognitive dimension
of language, which  makes it far more flexible  than species behavior.
Indeed, nothing similar to  multilinguism, diglossia or the appearance
of new  languages (pidgins and creoles) is  attested in non-linguistic
ecological systems. Modelling such type  of phenomena is still an open
challenge.

In sum,  as suggested by  Darwin, both languages and  ecosystems share
some  of  their  crucial  features.   These  would  include  spreading
dynamics, the presence of dramatic  thresholds or the role of space in
favouring heterogeneity.   In the language  context, this space-driven
enrichment can be interpreted in  other ways than physical space, such
as social distance. It is  also true, however, that a close inspection
of  both  systems  reveals   some  no  less  interesting  differences,
particularly  those  related  to  the flexibility  of  individuals  in
acquiring  several  languages or  the  social,  cultural or  political
factors  that constantly  interfere in  linguistic  phenomena.  Future
efforts towards  a theory of language change  might help understanding
our origins  as a complex, social  species and the  future of language
diversity.

\vspace{0.1 cm}

{\bf Acknowledgments}

\vspace{0.1 cm}

We thank  Guy Montag and  the members of  the Complex Systems  Lab for
useful  discussions. This  work  has been  supported  by NWO  research
project  Dependency  in  Universal  Grammar,  the  Spanish  MCIN  {\em
  Theoretical Linguistics}  2009SGR1079 (JF), the  James S.  McDonnell
Foundation (BCM) and by Santa Fe Institute (RS).

\begin{thereferences}

\item
Adami,  C.   1998.   {\em   Introduction  to  Artificial  Life}.   New
York:Springer.

\item
Arthur, B.  1994.  {\em Increasing returns and path  dependence in the
  economy}.  Michigan U. Press. Michigan.

\item
Atkinson,  Q.D.,  Meade, A.,  Venditti,  C.,  Greenhill, S.J.,  Pagel,
M.  2008.  Languages  Evolve  in Punctuational  Bursts. Science,  319,
588-588.

\item
Baronchelli, A., Felici, M., Caglioti, C., Loreto, V. and Steels, L. 2006. 
Sharp transition towards shared vocabularies in multi-agent systems. 
J. Stat. Mech. P06014.

\item
Baronchelli, A. Loreto, V. and Steels, L. 2008. 
In-depth analysis of the Naming Game dynamics: the homogeneous mixing case. 
Int. J. of Mod. Phys. C 19, 785-801.

\item
Bascompte, J. and Sol\'e, R. V. 2000. Rethinking Complexity: Modelling
Spatiotemporal Dynamics in Ecology. Trends Ecol Evol. 10, 361-366.

\item
Baxter,  G. J.   Blythe, R.  A. Croft,  W.  and  W.  McKane, A. J.  2006.
 Utterance  selection  model of  language  change.  Phys.  Rev. E  73,
 046118.

\item
Benedetto, D., Caglioti, E., Loreto, V. 2002. 
Language trees and zipping. Phys. Rev. Lett. 88, 048702. 

\item
Benett, C.H., Li, M. and Ma, B. 2003. Chain letters and evolutionary histories. 
Sci. Am. June 76-81.

\item
Bickerton, D.  1990. {\em Language  and Species}.  Chicago:  Chicago U
Press.

\item
Brighton, H. Smith, K. and Kirby, S. 2005. Language as an evolutionary
system. Phys. Life Rev. 2, 177-226.

\item
Buchanan,  M. 2003. {\em  Nexus: Small  Worlds and  the Groundbreaking
  Theory of Networks}. Norton and Co, New York, 2003.

\item
Cangelosi,  A.  and  Parisi, D.   1998.  Emergence  of language  in an
evolving population of neural networks.  Connection Science 10, 83-97.

\item
Cangelosi, A. 2001. Evolution of communication and language using 
signals, symbols, and words. IEEE Trans. Evol. Comp. 5, 93-101.

\item
Case, T. J. 1990.  Invasion resistance arises in strongly interacting 
species-rich model competition communities. 
Proc. Natl. Acad. Sci. USA 87, 9610-9614.

\item
Case, T. J. 1991.Invasion resistance, species build-up and community collapse 
in metapopulation models with interspecies competition. 
Biol. J. Linn. Soc. 42, 239-266.

\item
Case, T. J. 1999. {\em An Illustrated Guide to Theoretical Ecology}.  Oxford
U Press. New York.

\item
Cattuto,  C., Barrat,  A., Baldassarri,  A., Schehr,  G.,  and Loreto,
V.       2009.         Collective       dynamics       of       social
annotation. Proc. Natl. Acad. Sci. USA 106, 10511-10515.

\item
Castell\'o X., Eguiluz V. and San Miguel, M. 2006. Ordering dynamics with two
non-excluding options: bilingualism in language competition. New J. Phys. 8, 308.

\item
Cavalli-Sforza, L. L., Piazza, A., Menozzi, P., and Mountain, J. 1988. 
Reconstruction of human evolution: Bringing together genetic, 
archaeological, and linguistic data. Proc. Natl. Acad. Sci., USA 85, 6002-6006. 

\item
Cavalli-Sforza, L. 2002. {\em  Genes, people and language}. California
U. Press. Berkeley CA.

\item
Chapel, L., Castell\'o, X., Bernard, C., Deffuant, G., Eguiluz V., Martin, S. 
and San Miguel, M. 2010. 
Viability and Resilience of Languages in Competition. PLoS ONE 5, e8681.

\item
Chater, N., Reali, F. and Christiansen, M. H. 2009. Restrictions on biological 
adaptation in language evolution.  Proc. Natl. Acad. Sci., USA 106, 1015-1020.

\item
Chantrell,  G. 2002. {\em  The Oxford  dictionary of  word histories}.
Oxford U. Press. Oxford.

\item
Christiansen, M. H.,  and Chater, N. 2008.  Language  as shaped by the
brain. Behav Brain Sci. 31, 489-509.

\item
Crystal, D. 2000. {\em Language death}. Cambridge U. Press, UK.

\item
Dalby, A. 2003. Language in danger. Columbia U Press, New York.

\item
Dall'Asta, L., Baronchelli A, Barrat A, Loreto V. 2006. 
Nonequilibrium dynamics of language games on complex networks. 
Phys. Rev. E74, 036105.

\item
Darwin, C. 1871. {\em The descent of man, and selection in relation to
  sex}. John Murray, London. pp. 450.

\item
de  Oliveira,  V.  M.,  Gomes,  M.  A. F.,  and  Tsang,  I.  R.  2006.
Theoretical  model  for the  evolution  of  the linguistic  diversity.
Physica A 361, 361-370.

\item
de Oliveira, P., Stauffer, D., Wichmann, S., and Moss de Oliveira, S. 
2008. A computer simulation of language families. J. Linguistics 44, 659-675.

\item
Deacon, T.  W.  1997.  {\em  The Symbolic Species: The Co-evolution of
  language and brain}.  Norton, New York.

\item
DeGraff, M. 1999. {\em Language creation and language change: 
Creolization, diachrony and development}. MIT Press, Cambridge MA.

\item
Diamond, J. 1997. The language steamrollers. Nature 389, 544-546.

\item
Dieckmann,  U. Law,  R. and  Metz, J.  A. J.,  editors 2000.  {\em The
  geometry    of   ecological   interactions:    simplifying   spatial
  complexity}. Cambridge U. Press, Cambridge UK.

\item
Domingo,  E., Holland,  J. J.,  Biebricher,  C. and  Eigen, M  (1995),
Quasispecies: the  concept and the word, in:  {\em Molecular Evolution
  of  the Viruses},  (A. Gibbs,  C.  Calisher and  F. Garcia-  Arenal,
editors) Cambridge U. Press, Cambridge

\item
Eigen,   M.  McCaskill,   and   Schuster,  P.   1987.  The   Molecular
Quasispecies.  Adv. Chem. Phys. 75, 149-263

\item
Fishman, J. A. 1991. {\em Reversing language shift: Theoretical and empirical 
foundations of assistance to threatened languages}. Clevedon, UK.

\item
Fishman, J. A. (ed.) 2001. {\em Can Threatened Languages Be Saved? 
Reversing Language Shift, Revisited: A 21st Century Perspective}. Clevedon, UK. 

\item
Floreano, D., Mitri, S., Magnenat, S., and Keller, L. 2007. 
Evolutionary conditions for the emergence of communication in 
robots. Curr. Biol. 17, 514-519.

\item
Gaston,  K.   J.  2005.  Biodiversity  and   extinction:  species  and
people. Progr. Phys. Geography 29, 239–247.

\item
Gomes,  M. A.  F.  et al.  1999.  Scaling  relations for  diversity of
languages.  Physica A271: 489-495.

\item
Gong, T.,  Minett, J.  W., and Wang,  W. S-Y. 2008.   Exploring social
structure  effect  on  language  evolution based  on  a  computational
model. Connection Science 20, 135-153.

\item
Graddol, D. 2004. The future of language. Science 303, 1329 - 1331.

\item
Greenberg,  J.   (1963) Some  Universals  of  Grammar with  particular
reference  to  the  order  of  meaningful  elements.   In:  Greenberg,
J. H. (editor) {\em Universals of Language} MIT Press. London.

\item
Hauser, M.D.,  Chomsky, N., and Fitch,  W.  T.  2002.   The faculty of
language: what is  it, who has it and how did  it evolve? Science 298. 
1569-1579.

\item
Hawkins, J. and Gell-Mann, M.  1992. {\em The evolution of human languages}. 
Westview Press.

\item
Howe CJ, Barbrook AC, Spencer M, Robinson P, Bordalejo B, Mooney LR. 2001. 
Manuscript evolution. Trends Genet. 17, 147-52. 

\item
Kaplan, D. and Glass, L. 1995. {\em Understanding nonlinear dynamics}. 
Springer, New York.

\item
Ke,    J.,     Minnett,    J.     W.,    Au,    C.-P.     and    Wang,
W.  S-Y. 2002.  Self-organization and  selection in  the  emergence of
vocabulary. Complexity 7, 41-54.

\item
Ke, J.,  Gong, T. and Wang,  W. S-Y. 2008. Language  change and social
networks.  Comm. Comput. Phys. 2, 935-949.

\item
Kegl, J., Senghas, A and  Coppola, M. 1999.  Creation through Contact:
Language  Sign Emergence and  Sign Language  Change in  Nicaragua. In:
DeGraff,  Michel.    {\em  Language  Creation   and  Language  Change.
  Creolization,  Diachrony, and  Development}.  MIT  Press, Cambridge,
Mass.

\item
Kirby, S. 2002. Natural language from artificial life. 
Artificial Life 8, 185-215.

\item
Kosmidis,  K.,  Halley,  J.  M.,  and  Argyrakis,  P.  2005.  Language
evolution  and population  dynamics  in a  system  of two  interacting
species.  Physica A 353, 595-612.

\item
Kosmidis,  K., Kalampokis,  A., and  Argyrakis, P.  2006.  Statistical
Mechanical Approach to Human Language.  Physica A 366, 495-502.

\item
Lansing, J. S. et al., 2007. Coevolution of languages and genes on the
island of  Sumba, eastern Indonesia.  Proc. Natl. Acad. Sci.  USA 104,
16022-12026.

\item
Levins, R. 1968. {\em Evolution in changing environments}. 
Princeton U. Press. Princeton, NJ.

\item
Lieberman,  E.,  Michel,  J-B.,  Jackson,  J., Tang,  T.,  and  Nowak,
M.    A.   2007.    Quantifying    the   evolutionary    dynamics   of
language. Nature, 449, 713-716.

\item
Lipson, H. 2007. Evolutionary robotics: emergence of communication. 
Curr. Biol. 17, R330-R332.

\item
Liu, R-R., Jia, C-X., Yang, H-X., and Wang, B-H. 2009. 
Naming game on small-world networks with geographical effects. 
Physica A 388, 3615-3620.

\item
Loreto, V. and Steels, L. 2007. Emergence of language. Nature Phys. 3, 1-2.

\item
Lu, Q., Korniss, G., and Szymanski, B. K. 2008. Naming games in 
two-dimensional and small-world-connected 
random geometric networks. Physical Review E, 77(1).

\item
Mace R. and Pagel, M. 1995.   A Latitudinal Gradient in the Density of
Human Languages in North America.  Proc. R. Soc. Lond. B 261, 117-121.

\item
McWhorter,  J, 2001. {\em  The power  of Babel:  a natural  history of
  language}.  Harper and Collins, New York.

\item
Miller, G. A., 1991.  {\em The Science of Words}.  Scientific American
Library, New York: Freeman.

\item
Minett, J. W. and Wang, W. S-Y. 2008. Modelling endangered languages: 
The effects of bilingualism and social structure. Lingua, 118, 19-45.

\item
Mira, J. and A.  Paredes 2005. Interlinguistic simulation and language
death dynamics.  Europhys. Lett. 69, 1031-1034.

\item
Montoya,  J. M.,  Pimm,  S. L.,  and  Sol\'e, R.  V. 2006.  Ecological
networks and their fragility. Nature, 442, 259–264.

\item
Moore,  J.  L.,  L.  Manne,  T.  Brooks, N.  D.  Burgess,  R.  Davies,
C.  Rahbek, P.  Williams,and A.  Balmford. 2002.  The  distribution of
cultural and biological diversity in Africa. Proc. Royal Soc. London B
269, 1645–1653.

\item
Mufwene, S. 2004. Language birth and death. 
Annu. Rev. Anthropol. 33, 201-222.

\item
Murray, J. 1989. {\em Mathematical Biology}. Springer, New York.

\item
Nettle,  D. 1998. Explaining global patterns of language diversity. 
J. Antrop. Archeol. 17, 354-374.

\item
Nettle,  D. 1999a. {\em Linguistic diversity}. Oxford U. Press. Oxford.

\item
Nettle,  D. 1999b. Using social impact theory to simulate language 
change. Lingua 108, 95-117.

\item
Nettle,  D. 1999c. Is the rate of linguistic change constant? 
Lingua 108, 119-136.

\item
Nettle,  D.  and  Romaine,   S.  2002.  {\em  Vanishing  Voices.   The
  Extinction of the World's Languages}. Oxford U Press. Oxford UK.

\item
Niyogi, P. (2006)
{\em The Computational Nature of Language Learning and Evolution}. 
MIT Press. Cambridge, MA.

\item
Nolfi, S. and Mirolli, M. 2010. {\em 
Evolving Communication in Embodied Agents: Assessment and Open Challenges}. 
Springer, Berlin.

\item
Nowak,  M.   A.,  Krakauer,  D.   1999.  The  evolution  of  language.
Proc. Natl. Acad. Sci. USA 96, 8028-8033.

\item
Nowak M.A., Plotkin J.B., Jansen V. 2000. 
The evolution of syntactic communication. Nature 404, 495-498.

\item
Pagel, M. and Mace, R. 2004. The cultural wealth of nations. 
Nature 428, 275-278. 

\item
Pagel, M. 2009. Human language as a culturally transmitted 
replicator. Nature Rev. Genet. 10, 405-415.

\item
Parisi, D. 1997. An Artificial Life Approach to Language. 
Brain Lang. 59, 121-146.

\item
Patriarca, M.  and Leppanen,  T. 2004. Modeling  language competition.
Physica A 338, 296-299.

\item
Patriarca,  M.  and  Heinsalu,  E.  2008. Influence  of  geography  on
language competition. Physica A 388, 174-186.
 
\item
Pinker, S. 2000. Survival of the clearest. Nature 404, 441-442.

\item
Puglisi, A.,  Baronchelli, A, and  Loreto, V. 2008. Cultural  route to
the  emergence of linguistic  categories. Proc.  Natl. Acad.  Sci. USA
105, 7936-7940.

\item
Rosenzweig, M. L.   1995.  {\em Species Diversity in  Space and Time}.
Cambridge: Cambridge U. Press.

\item
Sandler,  W.,  Meir,  I.,  Padden,  C., and  Aronoff,  M.  2005.   The
emergence  of  grammar:  Systematic   structure  in  a  new  language.
Proc. Natl. Acad. Sci. USA 102, 2661-2665.

\item
Schulze C, Stauffer D, Wichmann S. 2008.  Birth, survival and death of
languages  by  Monte  Carlo  simulation.   Commun.  Comput.  Phys.  3,
271–294.

\item
Shen , Z-W. 1997. Exploring the dynamic aspect of sound change.
J. Chinese Linguist. Monograph Series Number 11.

\item
Silva, E. J. S. and de Oliveira, V. M. 2008. Evolution of the linguistic 
diversity on correlated landscapes. Physica A 387, 5597-5601.

\item
Smith,  K.  2002.  The   cultural  evolution  of  communication  in  a
population of neural networks. Conn. Sci. 14, 65-84.

\item
Sol\'e,  R. V.,  Bascompte,  J.  and Valls,  J.  1993.  Stability  and
Complexity    in    Spatially    Extended   Two-species    Competition
J. Theor. Biol. 159, 469-480.

\item
Sol\'e, R.  V., and  Goodwin, B.   C. 2001.  {\em  Signs of  Life: how
  complexity pervades biology}. New York: Basic Books.

\item
Sol\'e, R. V., Alonso, D. and McKane, A. 2002. 
Self-organized instability in complex ecosystems. Phil. Trans. Royal Soc. 
B 357, 667-681.

\item
Sol\'e,  R. V.,  Bascompte,  J.  2006. {\em Self-organization in complex 
ecosystems}. Princeton U. Press, Princeton.

\item
Stauffer,  D.  and  Schulze,  C.  2005.  Microscopic  and  macroscopic
simulation  of  competition  between  languages. Phys.  Life  Rev.  2,
89-116.

\item
Stauffer, D., Schulze, C., Lima,  F. W. S., Wichmann, S., and Solomon,
S. 2006.   Non-equilibrium and Irreversible  Simulation of Competition
among Languages.  Physica A 371, 719-724.

\item
Stauffer, D.,  Castello, X., Eguiluz, V.  N., and Miguel,  M. S. 2007.
Microscopic Abrams-Strogatz model  of language competition.  Physica A
374, 835-842.

\item
Steels, L. 2001. Language games for autonomous robots. IEEE Intel. Syst. 
October 16-22.

\item
Steels, L. and McIntyre, A. 1999. 
Spatially Distributed Naming Games. Adv. Complex Syst. 1, 301-323.

\item
Steels, L. 2003. Evolving grounded communication for robots. Trends in
Cognitive Science 7, 308-312.

\item
Steels, L. 2005. The  emergence and evolution of linguistic structure:
from lexical to grammatical communication systems.  Connection Science
17, 213-230.

\item
Sutherland,  W.   J.   2003.   Parallel  extinction  risk  and  global
distribution of languages and species. Nature 423, 276-279.

\item
Szamado, S. and Szathmary, E. 2006. Selective scenarios for the 
emergence of natural language. Trends Ecol. Evol. 21, 555-61. 

\item
Turing, A. 1952. The Chemical Basis of Morphogenesis. 
Phil. Trans. Royal Soc. B 237, 37-72.

\item
Wang, W. S-Y. 1969. Competing change as a cause of residue. Language 45, 9-25. 

\item
Wang, W. S-Y., Ke, J. and Minett, J. W. 2004. Computational studies of language 
evolution. Lang. Ling. Monograph Series B, 65–108.

\item
Wang,  W. S-Y.  and  Minett, J.  W.  2005. The  invasion of  language:
emergence, change and death. Trends Ecol. Evol. 20, 263-269.

\item
Whitfield,  J.  2008.   Across the  curious parallel  of  language and
species evolution.  PLOS Biol. 6, e186.

\item
Wray,   Martin   A.,   editor.    2002.   {\em   The   transition   to
  language}. Oxford: Oxford U. Press.

\item
Zanette, D. 2008. Analytical approach to bit string models of language
evolution. Int. J. Mod. Phys. C. 19, 569-581.

\end{thereferences}

\end{document}